\documentclass[
]{achemso}
\setkeys{acs}{articletitle=false} 
\usepackage{graphicx}% Include figure files
\usepackage{dcolumn}% Align table columns on decimal point
\usepackage{bm}% bold math
\usepackage{rotating}
\usepackage{tabularx}
\usepackage{amssymb}
\SectionNumbersOn
\usepackage[utf8]{inputenc}
\usepackage[T1]{fontenc}
\usepackage{mathptmx}
\usepackage{etoolbox}

\makeatletter
\def\@email#1#2{%
 \endgroup
 \patchcmd{\titleblock@produce}
  {\frontmatter@RRAPformat}
  {\frontmatter@RRAPformat{\produce@RRAP{*#1\href{mailto:#2}{#2}}}\frontmatter@RRAPformat}
  {}{}
}%
\makeatother

\author{Jonah Marks}
\author{Joseph Gomes}
\email{joe-gomes@uiowa.edu}
\affiliation[Iowa]
{Department of Chemical and Biochemical Engineering \\University of Iowa, Iowa City, United States}

\title[Efficient Transition State Searches by Freezing String Method with Graph Neural Network Potentials]{Efficient Transition State Searches by Freezing String Method with Graph Neural Network Potentials}

\abbreviations{MEP,TS,FSM,NNP,GNN,PES,XC}

\begin{document}

\begin{tocentry}

\includegraphics[width=\linewidth,height=\linewidth,keepaspectratio]{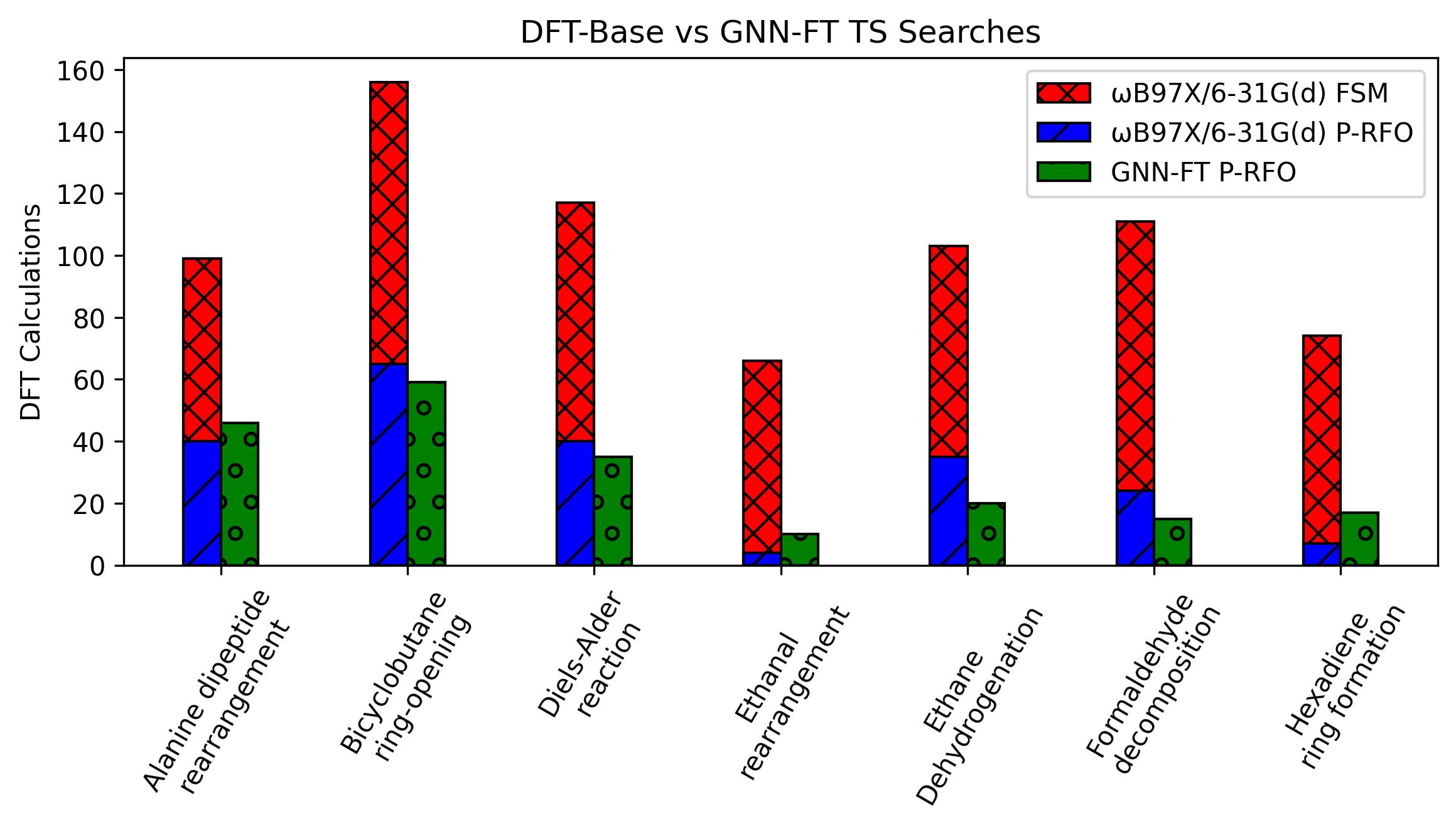}\\

Transition state (TS) searches are computationally intensive due to repeated evaluations of potential energy surfaces (PESs). We present a fine-tuned SchNet graph neural network (GNN) PES, pre-trained on ANI-1 and refined with reactant, product, and TS data, to accelerate TS searches via the Freezing String Method. Benchmarked on diverse reactions, our model successfully finds all TSs, and reduces computational costs by 72\% relative to full DFT searches, demonstrating GNNs as efficient surrogates for costly DFT PES evaluations.

\end{tocentry}

\begin{abstract}
Transition state (TS) searches are a critical bottleneck in computational studies of chemical reactivity, as accurately capturing complex phenomena like bond breaking and formation events requires repeated evaluations of expensive \textit{ab-initio} potential energy surfaces (PESs). While numerous algorithms have been developed to locate TSs efficiently, the computational cost of PES evaluations remains a key limitation. In this work, we develop and fine-tune a graph neural network (GNN) PES to accelerate TS searches for organic reactions. Our GNN of choice, SchNet, is first pre-trained on the ANI-1 dataset and subsequently fine-tuned on a small dataset of reactant, product, and TS structures. We integrate this GNN PES into the Freezing String Method (FSM), enabling rapid generation of TS guess geometries. Across a benchmark suite of chemically diverse reactions, our fine-tuned model (GNN-FT) achieves a 100\% success rate, locating the reference TSs in all cases while reducing the number of \textit{ab-initio} calculations by 72\% on average compared to conventional DFT-based FSM searches. Fine-tuning reduces GNN-FT errors by orders of magnitude for out-of-distribution cases such as non-covalent interactions, and improves TS-region predictions with comparatively little data. Analysis of transition state geometries and energy errors shows that GNN-FT captures PES along the reaction coordinate with sufficient accuracy to serve as a reliable DFT surrogate. These results demonstrate that modern GNN potentials, when properly trained, can significantly reduce the cost of TS searches and broaden the scope and size of systems considered in chemical reactivity studies.
\end{abstract}

\maketitle

\section{\label{sec:intro}Introduction}
Knowledge of transition states (TSs) is vital to accurately characterizing chemical reactions and predicting thermodynamic and kinetic rate parameters. However, TS are difficult to locate; they exist as first-order saddle points on the Born-Oppenheimer potential energy surface (PES) of atomic systems. To address the challenge of locating TS, a variety of algorithms and methods have been developed. TS search algorithms use first derivative and Hessian information to explore the PES and locate the TS, but this requires a suitable guess structure and a significant number of PES evaluations in series. 

TS search methods can be broadly categorized into surface walking\cite{cerjan1981finding,Banerjee1985,baker1986algorithm,wales1993locating,Heyden2005} methods and interpolation-based methods. Surface-walking algorithms maximize the largest negative eigenvalue of the Hessian matrix by moving uphill to locate the saddle point associated with that vibrational mode. In contrast, interpolation-based methods often split the search into two steps, leveraging two different algorithms. The first step of the interpolation-based approach is to use a non-local path finding algorithm to obtain a TS guess structure by efficiently stepping across the PES using only first derivative information. In the second step, the TS guess structure is used as the starting point for a surface walking algorithm to locate the exact TS. The success and efficiency of the overall search are heavily dependent on non-local path-finding algorithm; guess structures close to the saddle point of interest will quickly converge, minimizing the number of Hessian calculations required. However, guess structures that lie outside the basin of attraction can cause the surface walking algorithm to converge to an off-target critical point or fail to converge entirely. 

Development of reliable and efficient interpolation methods is a longstanding area of research. Early chain-of-states methods step across the PES to create a series of intermediate geometries connecting reactant and product\cite{Muller1979,Elber1987,ayala1997combined}. The string method\cite{Weinan2002} and the Nudged Elastic Band (NEB)\cite{Trygubenko2004,kolsbjerg2016automated,chu2003super,Henkelman2000,Henkelman2000CI,maragakis2002adaptive,ruttinger2022protocol} are double ended chain-of-states methods for estimating TS. These methods give information about the approximate minimum energy pathway and barrier height. Growing string methods construct an improved initial pathway from which a TS guess can be obtained by iteratively adding and optimizing intermediate structures\cite{Peters2004,Behn2011}. This allows the algorithm to avoid DFT calculations in non-physical areas of chemical space that initial interpolation paths may traverse. Reducing the calculation requirements of TS searches through algorithmic improvements to the reliability \cite{quapp2005growing,zimmerman2013reliable,zimmerman2015single,jafari2017reliable,suleimanov2015automated} and interpolation methods \cite{behn2011incorporating,Shaama2012,Zimmerman2013} for TS searches continues to be an area of active research.
These improvements have reduced the number of PES evaluations required to approximate the TS from thousands to under 100 electronic structure calculations\cite{marks2024incorporation}. 

% One issue with finding TSs is the efficiency of TS guess finding methods
Despite recent advancements in TS search algorithms, the underlying cost of the PES cannot be averted. Historically, potential energy functions trade level of detail for computational cost\cite{Behler2007,chmiela2018towards,zhang2018deep,anderson2019cormorant,park2021accurate}.
Inexpensive methods like atomic force fields, while efficient\cite{casewit1992application,stewart2007optimization}, often do not include sufficient detail to accurately describe chemical reactions. 
In some notable exceptions\cite{van2001reaxff,brenner2002second,Winetrout2024} reactive behavior can be parameterized into force fields; however, these methods are often fit to specific system types and are not broadly applicable when compared to quantum chemistry techniques like density functional theory (DFT). DFT scales cubically with respect to system size, placing a practical limit on system size that can be considered. Within DFT, exchange correlation (XC) functionals and basis set pairs determine the cost and accuracy of a simulation with more accurate XC functional/basis pairs incurring higher computational cost. One strategy to reduce TS search cost leverages this continuum by performing a TS search with a lower level of theory before re-calculating the final results at a higher level of theory\cite{Goodrow2008,goodrow2009transition}. Another method estimates the topology of the PES via interpolation to reduce the number of DFT calculations\cite{goodrow2010strategy}. 

In recent decades, machine learning (ML) models have proven to be powerful emulators of physics-based calculations (like DFT). Once trained, ML models require a fraction of the computational resources and exhibit favorable scaling with respect to system size. Early ML-based potentials, while innovative, failed to achieve chemical accuracy (error < 2~kcal/mol) across relevant areas of chemical space and were often fit to individual molecular systems, limiting broader applicability\cite{Behler2007,blank1995neural,Behler2007,handley2009optimal}. In contrast, modern ML models excel at approximating functions from large quantities of data. Several classes of neural network potentials (NNP) have emerged as computationally efficient approximators to expensive simulations like DFT: Behler-Parinello networks\cite{Behler2007,Smith2017ani1,Smith2019approaching}, graph neural networks\cite{Gilmer2017neural,schutt2017quantum,Schutt2018,unke2019physnet}, and equivariant networks\cite{gasteiger2020directional,batzner20223,musaelian2023learning}.

However, NNPs are heavily dependent on training data. Current datasets\cite{Ramakrishnan2014,Smith2017dataset} accurately describe minimum energy geometries and minorly perturbed structures. Training on these datasets results in models that achieve chemical accuracy on minimum or near-minimum energy structures\cite{Smith2017ani1,Schutt2018,Smith2019approaching}. This has enabled incorporation of ML potential energy functions into routine computational chemistry tasks like molecular dynamics\cite{Behler2007,noe2020machine,tkachenko2023neural} and geometry optimizations\cite{kulichenko2021rise,lan2023adsorbml}. However, accurate prediction of reaction parameters requires description of the TS, which lies in an area of chemical space inadequately sampled by current datasets. Consequently, models trained on current datasets are unable to accurately predict the potential energy and forces along reaction pathways prohibiting their use in reaction simulations\cite{Schreiner2022transition1x,zhang2024exploring}. Application of active learning and enhanced sampling methods has resulted in configurationally diverse datasets\cite{Smith2018less,vandermause2020fly,Smith2021automated,kulichenko2023uncertainty,zhang2024exploring}. Small datasets of reaction pathways\cite{Schreiner2022transition1x} and critical points of reactions \cite{Grambow2020dataset} have been published in recent years. Others have approached this problem by directly predicting reaction parameters, such as enthalpy of reaction, barrier heights, or TS geometry, from reactant and product structures or descriptors\cite{van20243dreact,Spiekermann2022,Heinen2021,grambow2020deep,Makos2021,Jackson2021,Pattanaik2020,stocker2020machine,meuwly2021machine,vanGerwen2022physics,Heid2021}. However, these methodologies are often limited in scope and are unable to reach chemical accuracy across broad benchmark test sets. Thus, despite the rapid advancement and incorporation of ML into computational chemistry, application of ML to TS searches has had limited success. 

Incorporation of ML potential energy functions into TS searches is a more universal approach. In TS searches, the process is broken into discrete steps
that use well understood techniques like geometry optimization, PES exploration algorithms, and saddle point search algorithms\cite{Baker1987,Heyden2005,peng1996using,bofill1994updated,henkelman1999dimer,mallikarjun2012automated,schlegel1982optimization,wales1993locating}. Substituting a ML PES for an \textit{ab-initio} PES in one or more of these tasks significantly reduces the computational cost of the TS search. This has been previously achieved through construction of approximate PES via Gaussian Process (GP) regression\cite{bartok2010gaussian,denzel2019gaussian,denzel2020hessian,xie2021bayesian}. While this methodology accelerates the overall TS search, the GP is fit on-the-fly to a given chemical system which requires \textit{ab-initio} calculations and cannot generalize to other systems; each TS search requires fitting a new GP from scratch. 

Two notable methods that incorporate ML PES into TS searches are CaTsunami and NeuralNEB. Both perform TS searches using pre-trained ML potentials and require no \textit{ab-initio} calculations to obtain a TS guess geometry. Both methods also use the NEB method with ML potential energy functions\cite{Schreiner2022NNEB,wander2024cattsunami}, significantly reducing the the computational cost of the NEB calculation. However, the additional error introduced by the ML potential energy function lowers the success rate of the TS searches. The authors of CaTSunami examine failed TS searches and find high root-mean-squared-error (RMSE) in relative potential energy predictions along the NEB path in these failed cases. 

In this work, we successfully demonstrate the incorporation of a graph neural network (GNN) PES into the freezing string method\cite{Behn2011} (FSM). Our NNP model of choice is the SchNet\cite{Schutt2018} GNN pre-trained using the ANI-1 dataset\cite{Smith2017dataset}. The pre-trained model accurately describes equilibrium and perturbed molecular structures within the training data distribution, although we find that it is not sufficient for TS searches as it poorly describes regions of chemical space where TSs exist. In the second training step, we improve our model by fine-tuning on a small dataset of reactant, TS, and product structures and energies based on the GDB7-20-TS dataset published by \citet{Grambow2020dataset}. We use the fine-tuned model to perform FSM calculations for a separate benchmark set of well-studied organic elementary reactions. We find that incorporation of a GNN potential energy function into the FSM significantly reduces the computational cost of TS searches associated with the non-local path finding step while maintaining a high TS search success rate. This work suggests that, with appropriate pre-training and fine-tuning, modern ML potential energy functions have reached a level of accuracy sufficient to be incorporated into routine computational chemistry tasks where computational cost is limiting factor.

\section{\label{sec:methods}Methods}

\subsection{\label{sec:datasets}Datasets}

In this work, we use the ANI-1 dataset\cite{Smith2017dataset} to pre-train a GNN potential energy function. The ANI-1 dataset contains approximately 20M off-equilibrium conformers of 57k molecules drawn from the GDB-11\cite{Fink2005,Fink2007} database of organic molecules. The molecules selected contain up to eight heavy (C-N-O) atoms with hydrogen atoms added to ensure neutral charge and are treated in their singlet electronic ground state. All structures and energies of the ANI-1 dataset are calculated at the $\omega$B97X/6-31G(d) level of theory. The structures included in the ANI-1 dataset result from normal mode sampling around equilibrium structures. Further details of the ANI-1 dataset can be found in the paper by Smith et al. 2017\cite{Smith2017dataset}.

Fine-tuning is performed with a second dataset, the GDB7-20-TS dataset, which contains structures and energies of the critical points (reactant, TS, and product) of approximately 12k gas phase elementary organic reactions\cite{Grambow2020dataset}. The structures and energies contained within GDB7-20-TS dataset are computed at the $\omega$B97X-D3/def2-TZVP level of theory. We re-optimize all structures and recompute potential energies at the $\omega$B97X/6-31G(d) level of theory. If a reactant, product, or TS fails the re-optimization, all structures associated with the reaction are removed from the fine-tuning set. Details and criteria for validating critical point structures can be found in Section \ref{sec:comp_details}. From the recomputed GDB7-20-TS dataset, 300 reactions are randomly sampled and reserved as a training validation set.

\subsection{\label{sec:GNNs} GNN Potential Energy Functions}
GNNs are a subset of neural networks that have had significant success within computational chemistry. To compose a graph, let $G=(V, E)$ denote a graph with node attributes $X_v$ for nodes $v \in V$. Given a set of graphs $\{G_1, ..., G_N\} \subseteq \mathcal{G}$ and their respective labels $\{y_1, ..., y_N\} \subseteq \mathcal{Y}$, the task of graph supervised learning is to learn a representation vector $h_G$ that serves to predict the label of an entire graph, $y_G = g(h_G)$. For chemical property prediction, $G$ is a molecular graph, where nodes represent atoms. Proximal atoms are connected by edges, and the label is the property to be predicted, such as potential energy.

In this work, we use the SchNet model \cite{Schutt2018} which learns representation vectors $h_v$ for every node $v \in \mathcal{G}$ through a general message passing scheme \cite{Gilmer2017neural}. In this strategy, the representation of a node is iteratively updated by aggregating the representations of its neighbors. After $k$ iterations of aggregation, the representation vector $h_v^{(k)}$ captures the structural information within its $k$-hop network neighborhood. The SchNet model employs a continuous convolution filter to weight the aggregated neighbor representations given the pairwise distance between nodes. The $k$-th layer of SchNet is described by equation \ref{eq:hv}
\begin{equation}
h_v^{(k)} = \sum_{u \in \mathcal{N}(v)} h_w^k \cdot W^k (r_w - r_v)
\label{eq:hv}
\end{equation}

where $h_v^{(k)}$ is the representation vector of node $v$ at the $k$-th layer, $\mathcal{N}(v)$ is the set neighbors of $v$, and $h_v^{(0)} = X_v$. $W^k$ is a filter generating neural network $W^l$: $\mathbb{R}^3$ $\rightarrow$ $\mathbb{R}^F$ that maps from atomic positions $r_w$ and $r_v$ to a value in the corresponding filter bank, weighting the representation vector based on neighbor distance. The total energy of a molecule is calculated as a sum of predictions of atomic contributions from vector representations from the final iteration K.
\begin{equation}
    \mathrm{Energy} = \sum_{N_{atoms}} \mathrm{MLP}(h_v^K)
\label{eq:2}
\end{equation}

\subsection{\label{sec:training_details} Training Details}
We use the SchNet model as implemented in PyTorch Geometric\cite{fey2019fast,Schutt2018}. To determine optimal model hyperparameters, we perform Bayesian optimization as implemented in the Ax adaptive experimentation program\cite{Snoek2012,Bakshy2018}. In Bayesian optimization, minimum and maximum values for each hyperparameter of interest are defined. An ensemble of trial models are trained, each with randomly sampled hyperparameter values within the pre-defined ranges. A Gaussian Process (GP) is then fit to the hyperparameter-validation performance, mapping from hyperparameter values to validation performance. Once the initial GP model is trained, future trials are selected through the expected improvement acquisition function, which balances exploring areas of high-performance and areas of high uncertainty in hyperparameter space. A new set of trials with hyperparameters selected by the GP is performed, and the resulting hyperparameter-validation performance results are added to the GP model. This process is repeated until a user-defined computational budget is reached. Further details of Bayesian Optimization, GP, and acquisition functions can be found elsewhere\cite{Snoek2012}. In this work, 96 trials are performed, with each model being trained for 50 epochs with validation early stopping on a single 2080TI NVIDIA GPU using the ANI-1 dataset using an 80/10/10 train/validatation/test split. The hyperparameters, their respective ranges, and optimized values are shown below in Table \ref{tab:bayeseopt}. Once all 96 trials were completed, we chose the best performing model based on validation mean absolute error (MAE) as our final model. 
%%table 2 here
\begin{table}
\caption{Hyperparameter ranges of the SchNet model considered during Bayesian optimization and final values of optimal pre-trained model, GNN-ANI1}
\label{tab:bayeseopt}
\begin{tabular}{p{4cm}p{4cm}p{4cm}}
    \hline
    Hyperparameter & Range & Final Value\\
    \hline
    Hidden Channels & 16-256 & 39\\
    Number of Filters & 16-256 & 111 \\
    Interaction Blocks & 2-14 & 12 \\
    Number of Gaussians & 16-128 & 105 \\
    Cutoff [{\AA}] & 2-12 & 8.0416 \\
    Learning Rate &$1^{-4}$-$2^{-3}$ & $2.94\times10^{-4}$\\
    Batch Size & - & 768 \\
    \hline
\end{tabular}
\end{table}

The optimal ANI-1 pre-trained model was fine-tuned on the GDB7-20-TS dataset re-optimized at the $\omega$B97X/6-31G(d) level of theory. During the fine-tuning, the model architecture is fixed and a limited hyperparameter search over the learning rate is performed on the fine-tuning set. All models are allowed to train for 100 epochs. To enable an \textit{a priori} selection of the final model, all candidate models are evaluated on the validation sets of both the ANI-1 and recomputed GDB7-20-TS datasets. The selected performing model results from training with a learning rate of 0.001 and achieves a mean-square-error (MSE) of 0.02~kcal/mol and 0.56~kcal/mol on the GDB7-20-TS and ANI-1 validation sets respectively. 

\subsection{\label{sec:TS_searches}Transition State Searches}

\begin{figure*}
\includegraphics[width=\textwidth,trim={0.5cm 2cm 0.5cm 1.6cm},clip]{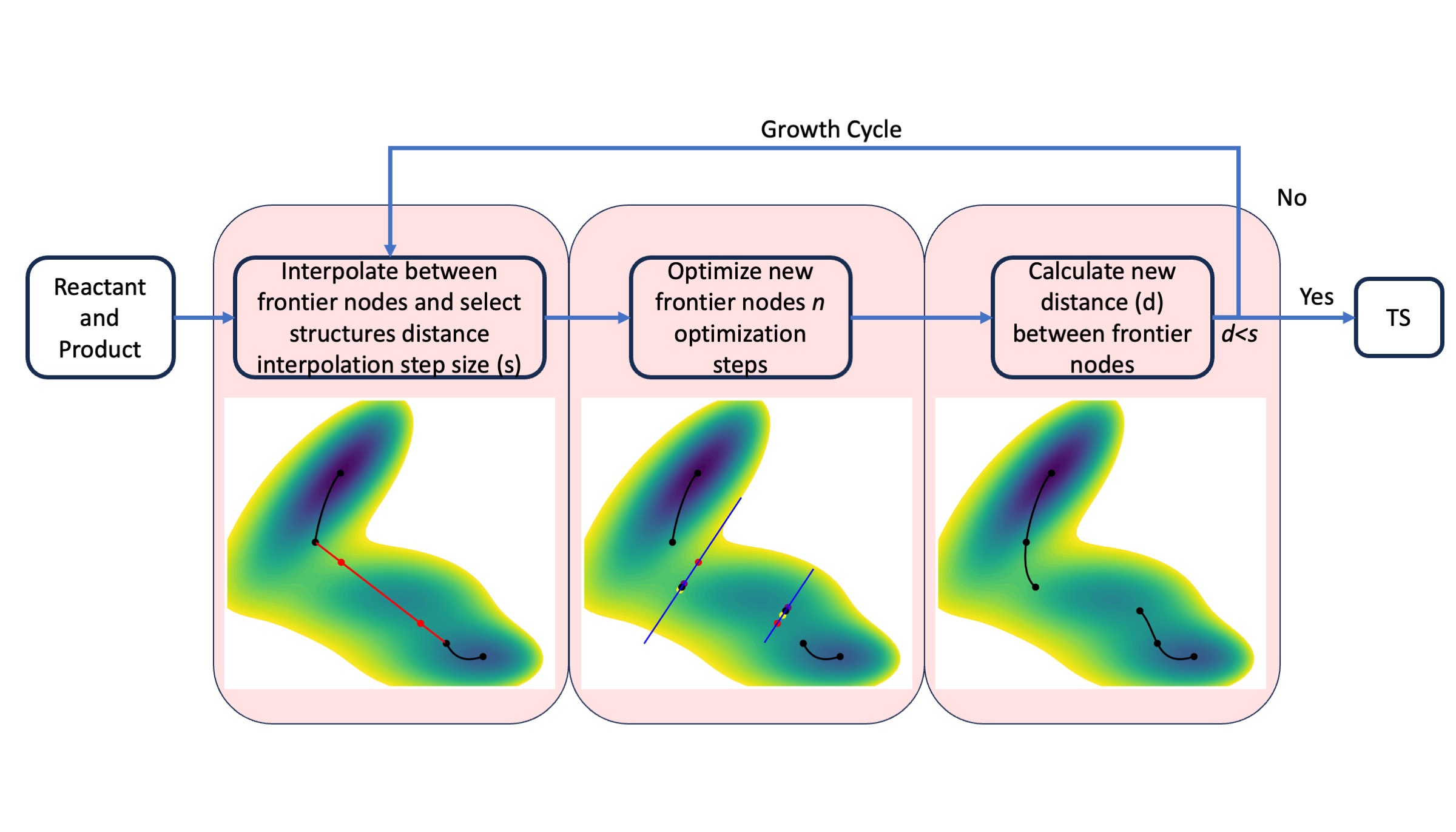}
\caption{Flow chart of the FSM with visualization of growth cycle steps on the M\"uller-Brown test potential. Interpolated paths and structures are shown in red, the node position after the first optimization cycle is shown in yellow, the position after the second optimization cycle is shown in purple, and the final (frozen) positions are shown in black.}
\label{fig:MB}
\end{figure*}

We use the FSM with internal coordinates interpolation in order to generate TS guess structures for refinement by local surface walking optimization. The details of adding new structures is illustrated in Figure \ref{fig:MB} using the M\"uller-Brown test potential\cite{Muller1979}. The approximate pathway takes a chain-of-states form consisting of intermediate structures along the reaction pathway. The approximate reaction pathway is iteratively built up by alternately adding interior nodes to reactant and product sides of a growing string. Interpolation is performed between the current innermost (frontier) nodes of the growing string. New reactant and product side structures are selected from the interpolated path based on a fixed step size $s$, and optimized on the hyperplane orthogonal to the current reaction path, ensuring the nodes lie close to the minimum energy pathway. After optimization, the geometries are frozen and will not change for the remainder of the calculation. This process is repeated until the reactant and product side strings meet. The highest energy node from the unified string is taken as the TS guess geometry and is refined to the exact TS via a local surface walking optimization algorithm. Details on the implementation of the FSM with internal coordinates interpolation are presented in Ref. \cite{marks2024incorporation}.

Four TS searches are performed and validated for each reaction in the test set. Two sets of TS search parameters are used. The baseline set of parameters were found to have the highest success rate across a chemically diverse benchmark set, while the efficient set was found to produce high quality TS guesses without significantly compromising success rate. In the baseline hyperparameter set, the minimum interpolation step size is chosen to be 1/18$^{th}$ the arc length distance between original reactant and product structures along the interpolated path, resulting in a FSM path with at minimum 18 nodes. Each optimization cycle is performed with at most three potential energy and force evaluations. When combined with the $\omega$B97X/6-31G(d) level of theory, we refer to this methodology as "DFT-Base". The efficient hyperparameter set is identical to the baseline set with the exception of the minimum number of nodes, which is reduced to nine, doubling the interpolation step size. When combined with the $\omega$B97X/6-31G(d) level of theory, this methodology is referred to as "DFT-Eff". Two searches are performed with the baseline hyperparameter set and use a pre-trained GNN potential energy function in the FSM portion of the TS search. One search is done with the model resulting from training on the ANI-1 dataset (GNN-ANI1). The other search is performed with the GNN potential energy functions trained on the ANI-1 dataset and fine-tuned on the GDB7-20-TS dataset (GNN-FT).

We calculate the cost of a TS search as the sum of the DFT calculations required by the FSM and local surface walking algorithm. When performing the FSM with a GNN potential energy function, it is assumed that it has negligible cost, as gradient calculations of neural networks are orders of magnitude less computationally expensive than \textit{ab-initio} calculations\cite{Smith2017ani1,Gilmer2017neural}. As a result, when the FSM is performed with a GNN PES, the entire cost of the TS search is incurred during the local surface walking algorithm. The TS guess structures from each FSM calculation are refined to their exact TS structure.

\subsection{\label{sec:comp_details}Computational Details}

Electronic structure calculations are performed to obtain optimized reactant and product geometries, compute quantum mechanical gradients and energies required by the FSM, to refine the TS guess to the exact transition state, and to perform intrinsic reaction coordinate (IRC) calculations. All electronic structure calculations were performed using the range separated hybrid generalized gradient approximation functional $\omega$B97X\cite{Chai2008} with the split-valence double-$\zeta$ polarized basis set, 6-31G(d)\cite{Ditchfield1971}, and a standard integration grid, SG-1\cite{gill1993standard}, as implemented in Q-Chem 6\cite{Epifanovsky2021}. This level of theory was chosen to maintain consistency with the ANI-1 pre-training dataset, thereby minimizing potential systematic errors. We report electronic energies without zero point correction. During reactant and product geometry optimization, energies were converged to 10$^{-6}$ Ha (Hartree), the maximum of the norm of the Cartesian gradient was converged to 10$^{-3}$ Ha bohr$^{-1}$, and final geometries were validated through vibrational frequency analysis, confirming the absence of imaginary frequencies. TS guess structures obtained from the FSM were refined to exact the transition states using the P-RFO method\cite{baker1986algorithm} with similar convergence criteria as for geometry optimization. P-RFO calculations that did not meet convergence criteria within 250 optimization steps were terminated. TS validation was performed by vibrational frequency calculations, ensuring the presence of a single imaginary frequency, and IRC calculations. In instances where IRC calculations prematurely terminated due to flat gradient regions, endpoint structures were further optimized using standard geometry optimization routines within Q-Chem, followed by energy profile analysis to confirm smooth energy descent. IRC calculations were performed using the predictor-corrector algorithm of \citet{schmidt1985intrinsic}. Details of the FSM software may be found elsewhere \cite{marks2024incorporation}, with additional modifications made to use GNN potentials. 

\section{\label{sec:ML}Machine Learning Results}

%IM Frequency and activation Energies calculated using QChem 1831 Runs
\begin{table}
    \caption{Thermodynamic parameters and imaginary frequency of the transition state of each reaction in the test set.}
    \label{tab:benchmarkset}
    \begin{tabularx}{\linewidth}{lXXX}
    \hline
    Reaction& $\Delta E^a$ & $E^{\ddag a}$ & Frequency$^b$ \\
        \hline
        Alanine Dipeptide Rearrangement & 1.1 & 6.9 & -54 \\
        C$_5$ $\leftrightarrow$ C$_{7AX}$ \\
        Bicyclobutane Ring-Opening & -13.9 & 57.8 & -283\\
        bicyclobutane $\leftrightarrow$ CH$_2$CHCHCH$_2$ \\
        Diels-Alder Reaction & -57.2 & 22.1 & -654\\
        CH$_2$CH$_2$ + CH$_2$CHCHCH$_2$ $\leftrightarrow$ cyclohexene \\
        Ethanal Proton Transfer & -16.0 & 59.0 & -2296\\
        CH$_2$CHOH $\leftrightarrow$ CH$_3$CHO \\
        Ethane Dehydrogenation & 47.0 & 135.2 & -2382\\
        C$_2$H$_6$ $\leftrightarrow$ C$_2$H$_4$ + H$_2$ \\
        Formaldehyde Decomposition & 10.3 & 95.8 & -2044\\
        H$_2$CO $\leftrightarrow$ H$_2$ + CO \\
        Hexadiene Ring Formation & 4.8 & 55.5 & -719\\
        cis,cis-2,4-hexadiene $\leftrightarrow$ 3,4-dimethylcyclobutene \\
        \hline
    \end{tabularx}
    \flushleft
    $^a {\mathrm{kcal/mol}}$\\
    $^b {\mathrm{cm}^{-1}}$\\
\end{table}

\subsection{\label{sec:Pre-training}Pre-training Results}
We first evaluate the best model from Bayesian optimization pre-training on the ANI-1 dataset, GNN-ANI1. The GNN-ANI1 model achieves a root-mean-square-error (RMSE) of 1.4~kcal/mol on the validation set of the ANI-1 dataset. To further evaluate the performance of the GNN-ANI1, we present a benchmark suite of reactions shown in Table \ref{tab:benchmarkset}. We evaluate the potential energies of reactant, TS, and product reference structures of each reaction and calculate the mean absolute error (MAE) with respect to $\omega$B97X/6-31G(d) values shown in Table \ref{tab:finetuning}. 
The GNN-ANI1 model achieves an MAE of 4.5, 34.5, and 11.3~kcal/mol on the reactants, products, and transition state structures respectively. Closer inspection of Table \ref{tab:finetuning} shows that the GNN-ANI1 achieves high accuracy on many structures in the benchmark set, but fails to reach chemical accuracy in several key areas. 

First, the GNN-ANI1 model struggles to accurately predict energetics of three- and four-membered rings. The product of hexadiene ring formation, 3,4-dimethylcyclobutene, is a four-membered ring containing a double bond. When predicting the potential energy of the product, the GNN-ANI1 has an error of 6.4~kcal/mol while the reactant, a linear alkene, has an error of 0.8~kcal/mol. Similarly, in the bicyclobutane ring opening reaction, the GNN-ANI1 model predicts a potential energy with an error of 16.8~kcal/mol for the reactant structure, while achieving an error of just 1.0~kcal/mol on the product. The reactant structure, bicyclobutane, is composed of two fused three-membered rings. The high intramolecular strain of such structures results in potential energies that deviate significantly from more common C-C bonding patterns.

Second, the products of the ethane dehydrogenation and formaldehyde decomposition reactions have the highest prediction errors across the entire benchmark set with errors of 81.0 and 148.9~kcal/mol respectively. These structures correspond to the dissociated states of the ethane dehydrogenation and formaldehyde decomposition reactions. Despite these being equilibrium structures, the GNN-ANI1 fails in this area of chemical space for two primary reasons. The ANI-1 dataset does not contain H-H bonding interactions, and does not contain intermolecular non-covalent interactions. Therefore, when the model predicts potential energies of dissociated H$_2$ systems, the model is far outside its training data distribution. The presence of H$_2$ appears to be a source of significant error; non-covalent interactions are present in the reactant state of the Diels-Alder reaction, but an error of only 7.2~kcal/mol is observed.
The final subset of structures, the transition states, is an area of chemical space that is not described by the ANI1 dataset and therefore is expected to consistently have high error. The lack of explicit H-H bonds and non-covalent interactions leads to lower prediction errors on TS structures than the product states of the ethane dehydrogenation and formaldehyde decomposition reactions. The GNN-ANI1 model does make reasonably accurate predictions for the transition states of the alanine dipeptide rearrangement and ethanal rearrangement reactions. These reactions correspond to rotation about dihedral angles (alanine dipeptide rearrangement) and a proton transfer (ethanal rearrangement). 

\begin{table}
    \centering
    \parbox{\textwidth}{ 
    \caption{Absolute error of GNN-ANI1 and GNN-FT energies, in kcal/mol, relative to the $\omega$B97X/6-31G(d) level of theory on reactants, transition states, and products from the reaction test set.}
    \label{tab:finetuning}
    \renewcommand{\arraystretch}{0.6}
    \centering
    \begin{tabular}{l ccc @{\hskip 12pt} ccc @{\hskip 12pt} c @{\hskip 12pt} c}
        \hline
        Reaction & \multicolumn{3}{c}{GNN-ANI1} & \multicolumn{3}{c}{GNN-FT} \\
        & Reactant & TS & Product & Reactant & TS & Product\\
        \hline
        Alanine Dipeptide Rearrangement & 3.7 & 0.3 & 3.1 & 1.5 & 0.3 & 3.0\\
        Bicyclobutane Ring-Opening & 16.8 & 20.0 & 1.0& 1.6 & 2.0 & 1.5\\
        Diels-Alder Reaction & 7.2 & 10.7 & 0.6& 0.6 & 4.3 & 0.2\\
        Ethanal Rearrangement & 1.7 & 1.9 & 0.4& 1.4 & 0.5 & 0.2\\
        Ethane Dehydrogenation & 0.5 & 22.5 & 81.0& 0.9 & 14.0 & 1.2\\
        Formaldehyde Decomposition & 0.5 & 19.0 & 148.9& 1.06 & 18.7 & 0.5\\
        Hexadiene Ring Formation & 0.8 & 4.8 & 6.4& 2.9 & 16.1 & 0.9\\
        \hline
    \end{tabular}
    }
\end{table}

\subsection{\label{sec:FT}Fine-Tuning Results}
We perform fine-tuning of the GNN-ANI1 model on the GDB7-20-TS dataset re-optimized to the $\omega$B97X/6-31G(d) level of theory. The outcome of fine-tuning is quantified by predicting the potential energy of the same set of benchmark structures shown in Table \ref{tab:finetuning}. The fine-tuning improves the model across all categories. The fine-tuned model (GNN-FT) has an MAE of 1.4, 1.1, and 8.0~kcal/mol on the reactants, products, and transition states respectively. 

We re-examine the four areas of poor performance before fine-tuning. In the product state of the hexadiene ring formation reaction, 3,4-dimethylcyclobutene, the error is reduced from 6.4 to 0.9~kcal/mol. Similarly, when predicting the potential energy of the bicyclobutane structure the error is reduced from 16.8 to 1.6~kcal/mol. A similar reduction in error is observed in the predictions of systems with non-covalent interactions. In the Diels-Alder reaction, prediction error of the reactant is reduced from 7.2 to 0.6~kcal/mol. For products of the ethane dehydrogenation and formaldehyde decomposition reactions the energy prediction error is reduced from 81.0 to 1.2~kcal/mol and 148.9 to 0.5~kcal/mol respectively. These substantial improvements indicate that the GNN-FT model is able to accurately describe non-covalent interactions in minimum energy configurations. Moreover, as GNN-FT model achieves chemical accuracy on the product states of both reactions demonstrating its capability to predict energetics of H-H bonding interactions. 

When predicting TS potential energies, the GNN-FT model has an average error of 8.0~kcal/mol, a 29\% reduction in MAE. This indicates that the fine tuning helps the model begin to make better predictions about TS geometries, but it still exceeds the chemical accuracy threshold, indicating the need for more data in this region of chemical space. The GNN potential shows an encouraging ability to learn energetics of new chemical phenomena such as H-H bonds, three- and four-membered rings, and intermolecular non-covalent interactions from relatively few data points when compared to the size of the pre-training set. This result highlights the possibility of fine-tuning of models for specific systems where little high-quality data is available. However, the GNN potential struggles to accurately predict energetics of TS phenomena like bond dissociation, formation, compression/stretching, etc. from the same number of examples. This limitation underscores the need for configurationally diverse quantum chemical datasets with balanced distributions that adequately describe these phenomena.

\section{\label{sec:TS_Results}Transition State Searches}
\subsection{\label{sec:Benchmark}Benchmark Set Results}

Our test set, shown in Table \ref{tab:benchmarkset}, is comprised of well-studied chemical reactions that are commonly used as test cases in TS search literature\cite{blanchard1966bicyclo,closs1968steric,dewar1975mindo,shevlin1988theoretical,mallikarjun2012automated}. This test set contains diverse reaction phenomena including isomerization, ring formation/opening, and bond dissociation/formation. First, TS searches are performed on each reaction using the DFT-Base parameters, giving a reference TS at the $\omega$B97X/6-31G(d) level of theory and a baseline computational cost. DFT-Base searches successfully find a TS for each reaction that is consistent with literature values, and is validated by vibrational frequency and IRC calculations. 

First, we perform FSM calculations using the GNN-ANI1 PES with subsequent refinement of the TS geometry at the $\omega$B97X/6-31G(d) level of theory. When performing the FSM with the GNN-ANI1 model, only five of the seven calculations yield a TS guess geometry that converges to a saddle point when refined. For both the formaldehyde decomposition and ethane dehydrogenation reactions, the guess geometry resulting from the GNN-ANI1 FSM either converges to a local minima structure (ethane dehydrogenation) or encounters a critical P-RFO error (formaldehyde decomposition). Both of these failed searches are discussed in greater detail in Section \ref{sec:failed_searches}. In the five successful cases, TS searches using the GNN-ANI1 average 37.6 DFT calculations while DFT-Base searches average 102.4 DFT calculations. This shows that when successful, using GNN PES to perform the FSM significantly reduces the computational cost of the TS search. 
%This point is no longer as good because there still is a large disparity
However, when the entire test set is compared, including failed runs, the GNN-ANI1 searches average 41.4 DFT calculations while DFT-Base runs average 103.7 DFT calculations. This underscores the importance of reliable TS searches. Failed searches not only fail to produce a TS, but also may consume significant computational resources before failing. Resubmission of FSM calculations and TS searches with new starting geometries would introduce further computational expenses.

%%%%%%%%%%%%%%%%%%%%%%%%%%%%%%%%%%%%%%%%%%%%%%%%%%%%%%%%%%%%

FSM calculations with the GNN-FT PES result in significantly better TS guess geometries than FSM calculations with the GNN-ANI1 model. GNN-FT calculations converge to the reference first-order saddle point in all searches. On average, GNN-FT-based TS searches require 28.9 DFT calculations to locate the exact TS while DFT-Base searches require 103.7 DFT calculations on average. Relative to searches with the GNN-ANI1 PES, a 30.2\% reduction in computational cost with a significant improvement in reliability as all GNN-FT searches find the reference saddle point. This large reduction in computational cost and increased reliability indicates that the GNN-FT model is able to more accurately describe the area of chemical space associated with transition states relative to the GNN-ANI1 model.

FSM hyperparameters play an important role in the cost and success of the TS search. TS searches are first performed using the baseline set of FSM hyperparameters, only varying the PES. This enables comparison of the effects of the PES on the success and cost of the overall TS search. However, the baseline hyperparameters are considered expensive parameters; small interpolation steps are taken, resulting in a dense final string. This results in accurate but expensive TS guesses. In a high-throughput screen it is much more realistic to use parameters that offer a balance between reliability and cost. We perform DFT searches using a more efficient set of FSM hyperparameters (DFT-Eff), which have twice the interpolation step size. As accuracy should be prioritized when selecting parameters, we do not perform new GNN-FT TS searches with the efficient hyperparameter set as the computational cost of the FSM is effectively negligible when performed with the GNN-FT PES. The efficient hyperparameter set results in a significant reduction in DFT calculations; DFT-Eff searches require 62.9 DFT calculations on average, a 39.3\% reduction compared to the results of the DFT-Base searches presented in Table \ref{tab:dft_cost}. The reduction in computational cost reduces the reliability of the TS search due to the coarser approximate reaction path. Despite the large reduction in computational cost, the DFT-Eff searches incur over twice the computational cost of the GNN-FT searches.

\begin{table}
    \centering
    \parbox{\textwidth}{
    \caption{DFT calculations required to locate the reference transition state using DFT-Base, DFT-Eff, GNN-ANI1, and GNN-FT parameters and models.}
    \label{tab:dft_cost}
    \renewcommand{\arraystretch}{0.6}
    \centering
    \begin{tabular}{l|ccc|ccc|c|c}
        \hline
        Reaction & \multicolumn{3}{c}{DFT-Base} & \multicolumn{3}{c}{DFT-Eff} & GNN-ANI1 & GNN-FT\\
        & FSM & TS & Total & FSM & TS & Total & Total & Total\\
        \hline
        Alanine Dipeptide Rearrangement & 59 & 40 & 99 & 24 & 97 & 121 & 67 & 46 \\
        Bicyclobutane Ring-Opening & 91 & 65 & 156 & 31 & 47 & 78 & 55 & 59 \\
        Diels-Alder Reaction & 77 & 40 & 117 & 20 & 41 & 61 & 32 & 35 \\
        Ethanal Rearrangement & 62 & 4 & 66 & 18 & 8 & 26 & 6 & 10 \\
        Ethane Dehydrogenation & 68 & 35 & 103 & 32 & 36 & 68 & 99* & 20 \\
        Formaldehyde Decomposition & 87 & 24 & 111 & 20 & 29 & 49 & 3* & 15 \\
        Hexadiene Ring Formation & 67 & 7 & 74 & 29 & 8 & 37 & 28 & 17\\
        \hline
    \end{tabular}
    }
\end{table}

\begin{table}
    \parbox{\textwidth}{
    \caption{Root mean squared displacement in Angstrom between TS guess structures obtained by FSM calculations compared to exact TS structure}
    \label{tab:RMSD}
    }
    \begin{tabular}{llrrrr}
    \hline
    Reaction & DFT-Base & DFT-Eff & GNN-ANI1 & GNN-FT \\
    \hline
    Alanine Dipeptide Rearrangement & 0.639 & 0.588 & 0.645 & 0.514 \\
    Bicyclobutane Ring-Opening & 0.463 & 0.451 & 0.443 & 0.504 \\
    Diels-Alder Reaction & 0.599 & 0.682 & 0.593 & 0.635 \\
    Ethanal Proton Transfer & 0.032 & 0.123 & 0.098 & 0.133 \\
    Ethane Dehydrogenation & 0.467 & 0.467 & 0.626 & 0.331 \\
    Formaldehyde Decomposition & 0.302 & 0.273 & 0.847 & 0.176 \\
    Hexadiene Ring Formation & 0.203 & 0.209 & 0.271 & 0.284 \\
    \hline
    Mean & 0.386 & 0.399 & 0.503 & 0.368 \\
    SEM & 0.083 & 0.077 & 0.096 & 0.071 \\
    \hline
    \end{tabular}
\end{table}

We examine the effects of replacing the $\omega$B97X/6-31G(d) PES with approximate GNN-ANI1 and GNN-FT PESs by considering the root-mean-square-displacement (RMSD) of TS guess geometry compared to the reference TS structure across our benchmark reaction set. The standard error of the mean (SEM) was used to quantify uncertainty. The results for each FSM methodology are summarized in Table \ref{tab:RMSD}. The TS guesses obtained from FSM calculations with the GNN-ANI1 algorithm have the highest average RMSD of 0.503~\AA~indicating that pretraining alone is insufficient to match DFT accuracy. TS guesses from the GNN-FT model have a substantially lower average RMSD of 0.368~\AA, a 26.8\% reduction relative to the GNN-ANI1. The GNN-FT model yields a lower mean RMSD than the DFT-Base and DFT-Eff references, however this difference is within statistical uncertainty and is not significant. Table \ref{tab:finetuning} shows that the GNN-FT model does not quantitatively approximate the energetics of the TS region of the $\omega$B97X/6-31G(d) PES. However, the low RMSD of TS guesses obtained using the GNN-FT suggests that the GNN-FT model approximates the reaction coordinate along the $\omega$B97X/6-31G(d) PES to a degree sufficient to serve as a DFT substitute.

\subsection{\label{sec:failed_searches}Failed TS Searches}
Ethane dehydrogenation has been demonstrated in prior literature\cite{mallikarjun2012automated} to be a difficult test case for FSM-based TS searches to identify the saddle point of interest. The TS guess structure resulting from the GNN-ANI1 FSM on the ethane dehydrogenation reaction has an initial RMSD of 0.626 and an initial imaginary frequency of -107$cm^{-1}$. Visual inspection of this structure show that it is similar to the fully dissociated product structure, and the frequency corresponds to rotation of the entire product complex. Inspection of the full final FSM path shows that the predicted highest energy structure is the product state which is expected given the error of the GNN-ANI1 model detailed in \ref{sec:Pre-training}. The reactants and products are explicitly excluded from the TS guess selection process. The selected structure is similar to the reactant state, with the H$_2$ molecule fully dissociated. The P-RFO algorithm attempts to maximize this imaginary mode, and after 99 optimization cycles the algorithm satisfies the gradient and energy change tolerances, and terminates.
In the case of formaldehyde decomposition an initial TS guess structure with three imaginary frequencies (-1193, -1026, and -116$cm^{-1}$) is used to start the P-RFO. After 3 optimization cycles the P-RFO algorithm fails to determine a valid optimization step, causing a fatal optimize error. This error can be resolved through a variety of methods but all require significant intervention and incur additional computational cost. The RMSD of all TS guess structures is shown in Table \ref{tab:RMSD}, the formaldehyde decomposition TS guess structure from the GNN-ANI1 FSM has an RMSD of 0.847~\AA, whereas guess structures from the FSM using the DFT-Base, DFT-Eff, and GNN-FT structures all are successfully refined to the reference TS and have RMSDs of 0.302, 0.273, and 0.176 respectively. The high RMSD, failed P-RFO optimization, and multiple strong imaginary frequencies all indicate that this is not a suitable TS guess. Because DFT-Base searches, which have identical parameters to GNN-ANI1 searches, are successful for both ethane dehydrogenation and formaldehyde decomposition, the failure is due to the change in PES used in the FSM. The GNN-ANI1 model performs particularly poorly across all areas of chemical space for systems with non-covalent interactions and H-H interactions, both of which are present in ethane dehydrogenation and formaldehyde decomposition.

\subsection{\label{sec:ala}Alanine Dipeptide}
\begin{figure}
\centering
    \includegraphics[width=\textwidth, trim=1.1cm 0 0 0, clip]{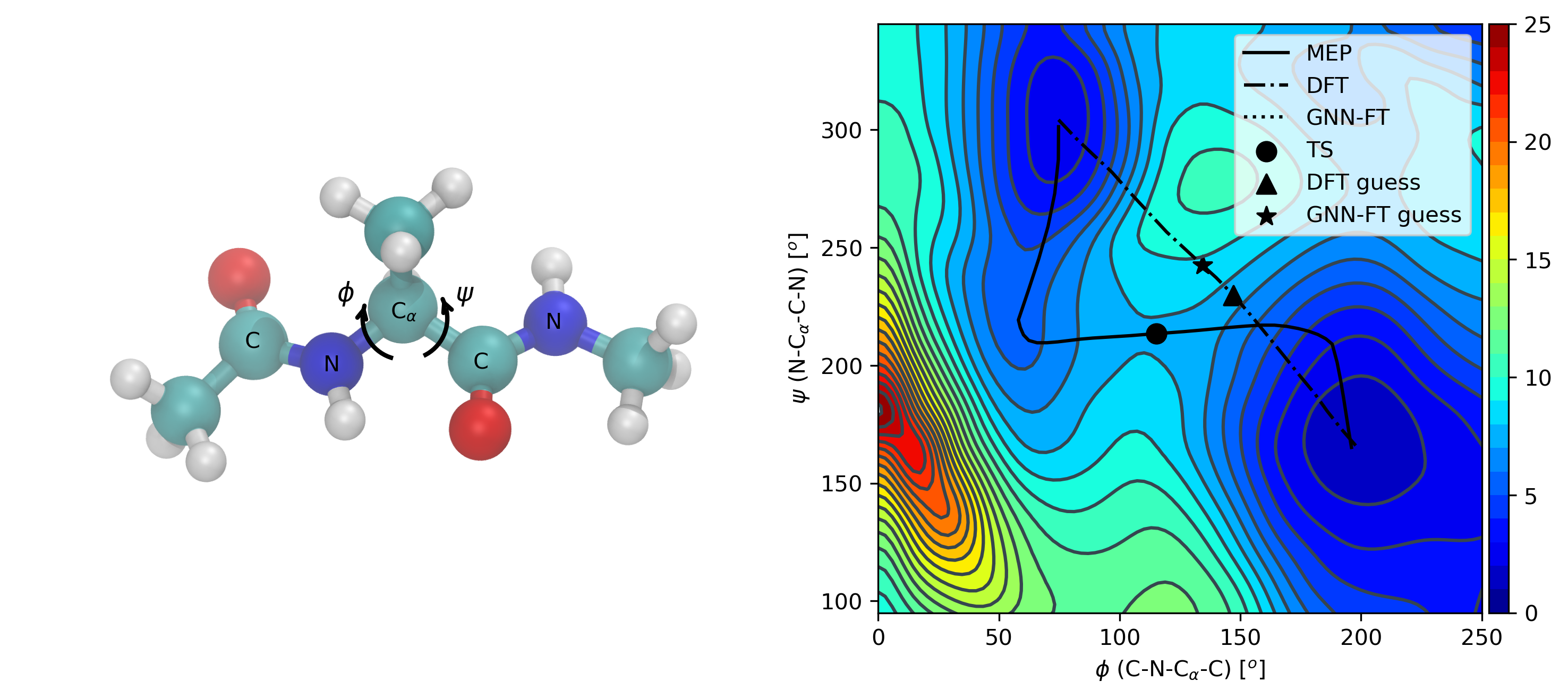}
    \caption{(Left) C$_5$ conformer of the alanine dipeptide system with Ramachandran angles labeled. (Right) Ramachandran angles along pathways taken by IRC, GNN-FT FSM, and $\omega$B97X/6-31G(d) FSM overlaid on the $\omega$B97X/6-31G(d) potential energy surface for the alanine dipeptide system.}
    \label{fig:alanine}
\end{figure}

Previous studies of the alanine dipeptide rearrangement reaction have examined the location of minimum energy structures (\citet{mironov2019systematic} and references therein), the minimum energy pathways of structural transitions\cite{Peters2004, behn2011incorporating, Shaama2012}, and the minimum free energy transition pathways\cite{vymetal2010metadynamics}. The alanine dipeptide exhibits multiple conformational minimum energy structures with well-characterized, relatively low transition barriers. It serves as a representative model of protein dynamics, making it an important test case for reaction path finding algorithms. This isomerization reaction involves the rotation about the $\phi$ (C-N-C$_\alpha$-C) and $\psi$ (N-C-C$_\alpha$-N) Ramachadran dihedral angles between the C$_5$ and C$_{7AX}$ conformations. In Figure \ref{fig:alanine}A we show the C$_5$ equilibrium conformation and indicate the Ramachadran angles. In Figure \ref{fig:alanine}B, the minimum energy pathway computed at the $\omega$B97X/6-31G(d) level of theory is projected onto the Ramachadran angles together with the approximate reaction pathways computed by the GNN-FT FSM and DFT-Base. The GNN-FT and DFT-Base use identical FSM hyperparameters and yield qualitatively similar pathways. This is reflected in the RMSD of the TS guess structures, the guess obtained from the GNN-FT has an RMSD of 0.514~\AA~while the guess from the small step size DFT-Base FSM has a RMSD of 0.639~\AA. The $\phi$ and $\psi$ dihedral coordinates undergo concerted rotation along the GNN-FT and DFT-Base pathways, while the IRC pathway shows a step-wise rotation of one dihedral coordinate followed by the other. A similar concerted FSM pathway has been reported by \citet{Behn2011} studying the identical reaction at the B3LYP/6-31G level of theory. Ultimately, both GNN-FT and DFT-Base TS guesses converge to the exact TS geometry when further refined.

\section{\label{sec:conclusion}Conclusion}

In this work, we demonstrate that the incorporation of an ML-based PES into the FSM TS search algorithm significantly reduces the computational cost associated with DFT calculations required for TS searches by nearly 50\% while maintaining a high success rate. This work shows that with appropriate training, ML PES representations are suitable for incorporation into TS searches and other routine computational chemistry tasks. We assess the accuracy and success rate of our approach based on a benchmark suite of several well-studied organic chemical reactions. In each test case, we successfully identify TS saddle point structures in fewer DFT calculations compared to the traditional approach. We show how fine-tuning our ML-based PES with the GDB7-20-TS dataset greatly improves model accuracy along reaction pathways, which ultimately leads to the success of the ML-based TS search algorithms. The reduction in computational cost can enable the use of such tools for the high-throughput study of reaction mechanisms and complete mapping of complex reaction networks. The incorporation of improved molecular representation learning algorithms \cite{lubbers2018hierarchical,batzner20223,gasteiger2020directional}, as well as datasets of configurationally diverse electronic structure calculations, is essential for further improvement of this work. 

\section*{Acknowledgments}
J.G acknowledges start up funding from The University of Iowa. This research was supported in part through computational resources provided by The University of Iowa.

\section*{Data Availability Statement}
The data that support the findings of this study are openly available in Zenodo at https://doi.org/10.5281/zenodo.14629429, reference number 14629429.

%\nocite{*}
\bibliography{main}% Produces the bibliography via BibTeX.

\providecommand{\latin}[1]{#1}
\makeatletter
\providecommand{\doi}
  {\begingroup\let\do\@makeother\dospecials
  \catcode`\{=1 \catcode`\}=2 \doi@aux}
\providecommand{\doi@aux}[1]{\endgroup\texttt{#1}}
\makeatother
\providecommand*\mcitethebibliography{\thebibliography}
\csname @ifundefined\endcsname{endmcitethebibliography}  {\let\endmcitethebibliography\endthebibliography}{}
\begin{mcitethebibliography}{105}
\providecommand*\natexlab[1]{#1}
\providecommand*\mciteSetBstSublistMode[1]{}
\providecommand*\mciteSetBstMaxWidthForm[2]{}
\providecommand*\mciteBstWouldAddEndPuncttrue
  {\def\EndOfBibitem{\unskip.}}
\providecommand*\mciteBstWouldAddEndPunctfalse
  {\let\EndOfBibitem\relax}
\providecommand*\mciteSetBstMidEndSepPunct[3]{}
\providecommand*\mciteSetBstSublistLabelBeginEnd[3]{}
\providecommand*\EndOfBibitem{}
\mciteSetBstSublistMode{f}
\mciteSetBstMaxWidthForm{subitem}{(\alph{mcitesubitemcount})}
\mciteSetBstSublistLabelBeginEnd
  {\mcitemaxwidthsubitemform\space}
  {\relax}
  {\relax}

\bibitem[Cerjan and Miller(1981)Cerjan, and Miller]{cerjan1981finding}
Cerjan,~C.~J.; Miller,~W.~H. \emph{J. Chem. Phys.} \textbf{1981}, \emph{75}, 2800--2806\relax
\mciteBstWouldAddEndPuncttrue
\mciteSetBstMidEndSepPunct{\mcitedefaultmidpunct}
{\mcitedefaultendpunct}{\mcitedefaultseppunct}\relax
\EndOfBibitem
\bibitem[Banerjee \latin{et~al.}(1985)Banerjee, Adams, Simons, and Shepard]{Banerjee1985}
Banerjee,~A.; Adams,~N.; Simons,~J.; Shepard,~R. \emph{J. Phys. Chem.} \textbf{1985}, \emph{89}, 52--57\relax
\mciteBstWouldAddEndPuncttrue
\mciteSetBstMidEndSepPunct{\mcitedefaultmidpunct}
{\mcitedefaultendpunct}{\mcitedefaultseppunct}\relax
\EndOfBibitem
\bibitem[Baker(1986)]{baker1986algorithm}
Baker,~J. \emph{J. Comput. Chem.} \textbf{1986}, \emph{7}, 385--395\relax
\mciteBstWouldAddEndPuncttrue
\mciteSetBstMidEndSepPunct{\mcitedefaultmidpunct}
{\mcitedefaultendpunct}{\mcitedefaultseppunct}\relax
\EndOfBibitem
\bibitem[Wales(1993)]{wales1993locating}
Wales,~D.~J. \emph{J. Chem. Soc. Faraday Trans.} \textbf{1993}, \emph{89}, 1305--1313\relax
\mciteBstWouldAddEndPuncttrue
\mciteSetBstMidEndSepPunct{\mcitedefaultmidpunct}
{\mcitedefaultendpunct}{\mcitedefaultseppunct}\relax
\EndOfBibitem
\bibitem[Heyden \latin{et~al.}(2005)Heyden, Bell, and Keil]{Heyden2005}
Heyden,~A.; Bell,~A.~T.; Keil,~F.~J. \emph{J. Chem. Phys.} \textbf{2005}, \emph{123}, 224101\relax
\mciteBstWouldAddEndPuncttrue
\mciteSetBstMidEndSepPunct{\mcitedefaultmidpunct}
{\mcitedefaultendpunct}{\mcitedefaultseppunct}\relax
\EndOfBibitem
\bibitem[Müller and Brown(1979)Müller, and Brown]{Muller1979}
Müller,~K.; Brown,~L.~D. \emph{Theor. Chim. Acta} \textbf{1979}, \emph{53}, 75--93\relax
\mciteBstWouldAddEndPuncttrue
\mciteSetBstMidEndSepPunct{\mcitedefaultmidpunct}
{\mcitedefaultendpunct}{\mcitedefaultseppunct}\relax
\EndOfBibitem
\bibitem[Elber and Karplus(1987)Elber, and Karplus]{Elber1987}
Elber,~R.; Karplus,~M. \emph{Chem. Phys. Lett.} \textbf{1987}, \emph{139}, 375--380\relax
\mciteBstWouldAddEndPuncttrue
\mciteSetBstMidEndSepPunct{\mcitedefaultmidpunct}
{\mcitedefaultendpunct}{\mcitedefaultseppunct}\relax
\EndOfBibitem
\bibitem[Ayala and Schlegel(1997)Ayala, and Schlegel]{ayala1997combined}
Ayala,~P.~Y.; Schlegel,~H.~B. \emph{J. Chem. Phys.} \textbf{1997}, \emph{107}, 375--384\relax
\mciteBstWouldAddEndPuncttrue
\mciteSetBstMidEndSepPunct{\mcitedefaultmidpunct}
{\mcitedefaultendpunct}{\mcitedefaultseppunct}\relax
\EndOfBibitem
\bibitem[Weinan \latin{et~al.}(2002)Weinan, Ren, and Vanden-Eijnden]{Weinan2002}
Weinan,~E.; Ren,~W.; Vanden-Eijnden,~E. \emph{Phys. Rev. B} \textbf{2002}, \emph{66}, 052301\relax
\mciteBstWouldAddEndPuncttrue
\mciteSetBstMidEndSepPunct{\mcitedefaultmidpunct}
{\mcitedefaultendpunct}{\mcitedefaultseppunct}\relax
\EndOfBibitem
\bibitem[Trygubenko and Wales(2004)Trygubenko, and Wales]{Trygubenko2004}
Trygubenko,~S.~A.; Wales,~D.~J. \emph{J. Chem. Phys.} \textbf{2004}, \emph{120}, 2082--2094\relax
\mciteBstWouldAddEndPuncttrue
\mciteSetBstMidEndSepPunct{\mcitedefaultmidpunct}
{\mcitedefaultendpunct}{\mcitedefaultseppunct}\relax
\EndOfBibitem
\bibitem[Kolsbjerg \latin{et~al.}(2016)Kolsbjerg, Groves, and Hammer]{kolsbjerg2016automated}
Kolsbjerg,~E.~L.; Groves,~M.~N.; Hammer,~B. \emph{J. Chem. Phys.} \textbf{2016}, \emph{145}, 094107\relax
\mciteBstWouldAddEndPuncttrue
\mciteSetBstMidEndSepPunct{\mcitedefaultmidpunct}
{\mcitedefaultendpunct}{\mcitedefaultseppunct}\relax
\EndOfBibitem
\bibitem[Chu \latin{et~al.}(2003)Chu, Trout, and Brooks]{chu2003super}
Chu,~J.-W.; Trout,~B.~L.; Brooks,~B.~R. \emph{J. Chem. Phys.} \textbf{2003}, \emph{119}, 12708--12717\relax
\mciteBstWouldAddEndPuncttrue
\mciteSetBstMidEndSepPunct{\mcitedefaultmidpunct}
{\mcitedefaultendpunct}{\mcitedefaultseppunct}\relax
\EndOfBibitem
\bibitem[Henkelman and Jónsson(2000)Henkelman, and Jónsson]{Henkelman2000}
Henkelman,~G.; Jónsson,~H. \emph{J. Chem. Phys.} \textbf{2000}, \emph{113}, 9978--9985\relax
\mciteBstWouldAddEndPuncttrue
\mciteSetBstMidEndSepPunct{\mcitedefaultmidpunct}
{\mcitedefaultendpunct}{\mcitedefaultseppunct}\relax
\EndOfBibitem
\bibitem[Henkelman \latin{et~al.}(2000)Henkelman, Uberuaga, and Jonsson]{Henkelman2000CI}
Henkelman,~G.; Uberuaga,~B.~P.; Jonsson,~H. \emph{J. Chem. Phys.} \textbf{2000}, \emph{113}, 9901--9904\relax
\mciteBstWouldAddEndPuncttrue
\mciteSetBstMidEndSepPunct{\mcitedefaultmidpunct}
{\mcitedefaultendpunct}{\mcitedefaultseppunct}\relax
\EndOfBibitem
\bibitem[Maragakis \latin{et~al.}(2002)Maragakis, Andreev, Brumer, Reichman, and Kaxiras]{maragakis2002adaptive}
Maragakis,~P.; Andreev,~S.~A.; Brumer,~Y.; Reichman,~D.~R.; Kaxiras,~E. \emph{J. Chem. Phys.} \textbf{2002}, \emph{117}, 4651--4658\relax
\mciteBstWouldAddEndPuncttrue
\mciteSetBstMidEndSepPunct{\mcitedefaultmidpunct}
{\mcitedefaultendpunct}{\mcitedefaultseppunct}\relax
\EndOfBibitem
\bibitem[Ruttinger \latin{et~al.}(2022)Ruttinger, Sharma, and Clancy]{ruttinger2022protocol}
Ruttinger,~A.~W.; Sharma,~D.; Clancy,~P. \emph{J. Chem. Theory Comput.} \textbf{2022}, \emph{18}, 2993--3005\relax
\mciteBstWouldAddEndPuncttrue
\mciteSetBstMidEndSepPunct{\mcitedefaultmidpunct}
{\mcitedefaultendpunct}{\mcitedefaultseppunct}\relax
\EndOfBibitem
\bibitem[Peters \latin{et~al.}(2004)Peters, Heyden, Bell, and Chakraborty]{Peters2004}
Peters,~B.; Heyden,~A.; Bell,~A.~T.; Chakraborty,~A. \emph{J. Chem. Phys.} \textbf{2004}, \emph{120}, 7877--7886\relax
\mciteBstWouldAddEndPuncttrue
\mciteSetBstMidEndSepPunct{\mcitedefaultmidpunct}
{\mcitedefaultendpunct}{\mcitedefaultseppunct}\relax
\EndOfBibitem
\bibitem[Behn \latin{et~al.}(2011)Behn, Zimmerman, Bell, and Head-Gordon]{Behn2011}
Behn,~A.; Zimmerman,~P.~M.; Bell,~A.~T.; Head-Gordon,~M. \emph{J. Chem. Phys.} \textbf{2011}, \emph{135}, 224108\relax
\mciteBstWouldAddEndPuncttrue
\mciteSetBstMidEndSepPunct{\mcitedefaultmidpunct}
{\mcitedefaultendpunct}{\mcitedefaultseppunct}\relax
\EndOfBibitem
\bibitem[Quapp(2005)]{quapp2005growing}
Quapp,~W. \emph{J. Chem. Phys.} \textbf{2005}, \emph{122}, 174106\relax
\mciteBstWouldAddEndPuncttrue
\mciteSetBstMidEndSepPunct{\mcitedefaultmidpunct}
{\mcitedefaultendpunct}{\mcitedefaultseppunct}\relax
\EndOfBibitem
\bibitem[Zimmerman(2013)]{zimmerman2013reliable}
Zimmerman,~P. \emph{J. Chem. Theory Comput.} \textbf{2013}, \emph{9}, 3043--3050\relax
\mciteBstWouldAddEndPuncttrue
\mciteSetBstMidEndSepPunct{\mcitedefaultmidpunct}
{\mcitedefaultendpunct}{\mcitedefaultseppunct}\relax
\EndOfBibitem
\bibitem[Zimmerman(2015)]{zimmerman2015single}
Zimmerman,~P.~M. \emph{J. Comput. Chem.} \textbf{2015}, \emph{36}, 601--611\relax
\mciteBstWouldAddEndPuncttrue
\mciteSetBstMidEndSepPunct{\mcitedefaultmidpunct}
{\mcitedefaultendpunct}{\mcitedefaultseppunct}\relax
\EndOfBibitem
\bibitem[Jafari and Zimmerman(2017)Jafari, and Zimmerman]{jafari2017reliable}
Jafari,~M.; Zimmerman,~P.~M. \emph{J. Comput. Chem.} \textbf{2017}, \emph{38}, 645--658\relax
\mciteBstWouldAddEndPuncttrue
\mciteSetBstMidEndSepPunct{\mcitedefaultmidpunct}
{\mcitedefaultendpunct}{\mcitedefaultseppunct}\relax
\EndOfBibitem
\bibitem[Suleimanov and Green(2015)Suleimanov, and Green]{suleimanov2015automated}
Suleimanov,~Y.~V.; Green,~W.~H. \emph{J. Chem. Theory Comput.} \textbf{2015}, \emph{11}, 4248--4259\relax
\mciteBstWouldAddEndPuncttrue
\mciteSetBstMidEndSepPunct{\mcitedefaultmidpunct}
{\mcitedefaultendpunct}{\mcitedefaultseppunct}\relax
\EndOfBibitem
\bibitem[Behn \latin{et~al.}(2011)Behn, Zimmerman, Bell, and Head-Gordon]{behn2011incorporating}
Behn,~A.; Zimmerman,~P.~M.; Bell,~A.~T.; Head-Gordon,~M. \emph{J. Chem. Theory Comput.} \textbf{2011}, \emph{7}, 4019--4025\relax
\mciteBstWouldAddEndPuncttrue
\mciteSetBstMidEndSepPunct{\mcitedefaultmidpunct}
{\mcitedefaultendpunct}{\mcitedefaultseppunct}\relax
\EndOfBibitem
\bibitem[Sharada \latin{et~al.}(2012)Sharada, Zimmerman, Bell, and Head-Gordon]{Shaama2012}
Sharada,~S.~M.; Zimmerman,~P.~M.; Bell,~A.~T.; Head-Gordon,~M. \emph{J. Chem. Theory Comput.} \textbf{2012}, \emph{8}, 5166--5174\relax
\mciteBstWouldAddEndPuncttrue
\mciteSetBstMidEndSepPunct{\mcitedefaultmidpunct}
{\mcitedefaultendpunct}{\mcitedefaultseppunct}\relax
\EndOfBibitem
\bibitem[Zimmerman(2013)]{Zimmerman2013}
Zimmerman,~P.~M. \emph{J. Chem. Phys.} \textbf{2013}, \emph{138}, 184102\relax
\mciteBstWouldAddEndPuncttrue
\mciteSetBstMidEndSepPunct{\mcitedefaultmidpunct}
{\mcitedefaultendpunct}{\mcitedefaultseppunct}\relax
\EndOfBibitem
\bibitem[Marks and Gomes(2024)Marks, and Gomes]{marks2024incorporation}
Marks,~J.; Gomes,~J. \emph{arXiv preprint arXiv:2407.09763} \textbf{2024}, \relax
\mciteBstWouldAddEndPunctfalse
\mciteSetBstMidEndSepPunct{\mcitedefaultmidpunct}
{}{\mcitedefaultseppunct}\relax
\EndOfBibitem
\bibitem[Behler and Parrinello(2007)Behler, and Parrinello]{Behler2007}
Behler,~J.; Parrinello,~M. \emph{Phys. Rev. Lett.} \textbf{2007}, \emph{98}, 146401\relax
\mciteBstWouldAddEndPuncttrue
\mciteSetBstMidEndSepPunct{\mcitedefaultmidpunct}
{\mcitedefaultendpunct}{\mcitedefaultseppunct}\relax
\EndOfBibitem
\bibitem[Chmiela \latin{et~al.}(2018)Chmiela, Sauceda, M{\"u}ller, and Tkatchenko]{chmiela2018towards}
Chmiela,~S.; Sauceda,~H.~E.; M{\"u}ller,~K.-R.; Tkatchenko,~A. \emph{Nat. Commun.} \textbf{2018}, \emph{9}, 3887\relax
\mciteBstWouldAddEndPuncttrue
\mciteSetBstMidEndSepPunct{\mcitedefaultmidpunct}
{\mcitedefaultendpunct}{\mcitedefaultseppunct}\relax
\EndOfBibitem
\bibitem[Zhang \latin{et~al.}(2018)Zhang, Han, Wang, Car, and E]{zhang2018deep}
Zhang,~L.; Han,~J.; Wang,~H.; Car,~R.; E,~W. \emph{Phys. Rev. Lett.} \textbf{2018}, \emph{120}, 143001\relax
\mciteBstWouldAddEndPuncttrue
\mciteSetBstMidEndSepPunct{\mcitedefaultmidpunct}
{\mcitedefaultendpunct}{\mcitedefaultseppunct}\relax
\EndOfBibitem
\bibitem[Anderson \latin{et~al.}(2019)Anderson, Hy, and Kondor]{anderson2019cormorant}
Anderson,~B.; Hy,~T.~S.; Kondor,~R. Cormorant: Covariant Molecular Neural Networks. Advances in Neural Information Processing Systems 32. Vancouver, Canada, 2019; pp 14510--14519\relax
\mciteBstWouldAddEndPuncttrue
\mciteSetBstMidEndSepPunct{\mcitedefaultmidpunct}
{\mcitedefaultendpunct}{\mcitedefaultseppunct}\relax
\EndOfBibitem
\bibitem[Park \latin{et~al.}(2021)Park, Kornbluth, Vandermause, Wolverton, Kozinsky, and Mailoa]{park2021accurate}
Park,~C.~W.; Kornbluth,~M.; Vandermause,~J.; Wolverton,~C.; Kozinsky,~B.; Mailoa,~J.~P. \emph{npj Comput. Mater.} \textbf{2021}, \emph{7}, 73\relax
\mciteBstWouldAddEndPuncttrue
\mciteSetBstMidEndSepPunct{\mcitedefaultmidpunct}
{\mcitedefaultendpunct}{\mcitedefaultseppunct}\relax
\EndOfBibitem
\bibitem[Casewit \latin{et~al.}(1992)Casewit, Colwell, and Rappe]{casewit1992application}
Casewit,~C.; Colwell,~K.; Rappe,~A. \emph{J. Am. Chem. Soc.} \textbf{1992}, \emph{114}, 10035--10046\relax
\mciteBstWouldAddEndPuncttrue
\mciteSetBstMidEndSepPunct{\mcitedefaultmidpunct}
{\mcitedefaultendpunct}{\mcitedefaultseppunct}\relax
\EndOfBibitem
\bibitem[Stewart(2007)]{stewart2007optimization}
Stewart,~J.~J. \emph{J. Mol. Model.} \textbf{2007}, \emph{13}, 1173--1213\relax
\mciteBstWouldAddEndPuncttrue
\mciteSetBstMidEndSepPunct{\mcitedefaultmidpunct}
{\mcitedefaultendpunct}{\mcitedefaultseppunct}\relax
\EndOfBibitem
\bibitem[Van~Duin \latin{et~al.}(2001)Van~Duin, Dasgupta, Lorant, and Goddard]{van2001reaxff}
Van~Duin,~A.~C.; Dasgupta,~S.; Lorant,~F.; Goddard,~W.~A. \emph{J. Phys. Chem. A} \textbf{2001}, \emph{105}, 9396--9409\relax
\mciteBstWouldAddEndPuncttrue
\mciteSetBstMidEndSepPunct{\mcitedefaultmidpunct}
{\mcitedefaultendpunct}{\mcitedefaultseppunct}\relax
\EndOfBibitem
\bibitem[Brenner \latin{et~al.}(2002)Brenner, Shenderova, Harrison, Stuart, Ni, and Sinnott]{brenner2002second}
Brenner,~D.~W.; Shenderova,~O.~A.; Harrison,~J.~A.; Stuart,~S.~J.; Ni,~B.; Sinnott,~S.~B. \emph{J. Phys.:Condens. Matter} \textbf{2002}, \emph{14}, 783\relax
\mciteBstWouldAddEndPuncttrue
\mciteSetBstMidEndSepPunct{\mcitedefaultmidpunct}
{\mcitedefaultendpunct}{\mcitedefaultseppunct}\relax
\EndOfBibitem
\bibitem[Winetrout \latin{et~al.}(2024)Winetrout, Kanhaiya, Kemppainen, in~‘t Veld, Sachdeva, Pandey, Damirchi, van Duin, Odegard, and Heinz]{Winetrout2024}
Winetrout,~J.~J.; Kanhaiya,~K.; Kemppainen,~J.; in~‘t Veld,~P.~J.; Sachdeva,~G.; Pandey,~R.; Damirchi,~B.; van Duin,~A.; Odegard,~G.~M.; Heinz,~H. \emph{Nat. Commun.} \textbf{2024}, \emph{15}\relax
\mciteBstWouldAddEndPuncttrue
\mciteSetBstMidEndSepPunct{\mcitedefaultmidpunct}
{\mcitedefaultendpunct}{\mcitedefaultseppunct}\relax
\EndOfBibitem
\bibitem[Goodrow \latin{et~al.}(2008)Goodrow, Bell, and Head-Gordon]{Goodrow2008}
Goodrow,~A.; Bell,~A.~T.; Head-Gordon,~M. \emph{J. Chem. Phys.} \textbf{2008}, \emph{129}, 174109\relax
\mciteBstWouldAddEndPuncttrue
\mciteSetBstMidEndSepPunct{\mcitedefaultmidpunct}
{\mcitedefaultendpunct}{\mcitedefaultseppunct}\relax
\EndOfBibitem
\bibitem[Goodrow \latin{et~al.}(2009)Goodrow, Bell, and Head-Gordon]{goodrow2009transition}
Goodrow,~A.; Bell,~A.~T.; Head-Gordon,~M. \emph{J. Chem. Phys.} \textbf{2009}, \emph{130}, 244108\relax
\mciteBstWouldAddEndPuncttrue
\mciteSetBstMidEndSepPunct{\mcitedefaultmidpunct}
{\mcitedefaultendpunct}{\mcitedefaultseppunct}\relax
\EndOfBibitem
\bibitem[Goodrow \latin{et~al.}(2010)Goodrow, Bell, and Head-Gordon]{goodrow2010strategy}
Goodrow,~A.; Bell,~A.~T.; Head-Gordon,~M. \emph{Chem. Phys. Lett.} \textbf{2010}, \emph{484}, 392--398\relax
\mciteBstWouldAddEndPuncttrue
\mciteSetBstMidEndSepPunct{\mcitedefaultmidpunct}
{\mcitedefaultendpunct}{\mcitedefaultseppunct}\relax
\EndOfBibitem
\bibitem[Blank \latin{et~al.}(1995)Blank, Brown, Calhoun, and Doren]{blank1995neural}
Blank,~T.~B.; Brown,~S.~D.; Calhoun,~A.~W.; Doren,~D.~J. \emph{J. Chem. Phys.} \textbf{1995}, \emph{103}, 4129--4137\relax
\mciteBstWouldAddEndPuncttrue
\mciteSetBstMidEndSepPunct{\mcitedefaultmidpunct}
{\mcitedefaultendpunct}{\mcitedefaultseppunct}\relax
\EndOfBibitem
\bibitem[Handley \latin{et~al.}(2009)Handley, Hawe, Kell, and Popelier]{handley2009optimal}
Handley,~C.~M.; Hawe,~G.~I.; Kell,~D.~B.; Popelier,~P.~L. \emph{Phys. Chem. Chem. Phys.} \textbf{2009}, \emph{11}, 6365--6376\relax
\mciteBstWouldAddEndPuncttrue
\mciteSetBstMidEndSepPunct{\mcitedefaultmidpunct}
{\mcitedefaultendpunct}{\mcitedefaultseppunct}\relax
\EndOfBibitem
\bibitem[Smith \latin{et~al.}(2017)Smith, Isayev, and Roitberg]{Smith2017ani1}
Smith,~J.~S.; Isayev,~O.; Roitberg,~A.~E. \emph{Chem. Sci.} \textbf{2017}, \emph{8}, 3192--3203\relax
\mciteBstWouldAddEndPuncttrue
\mciteSetBstMidEndSepPunct{\mcitedefaultmidpunct}
{\mcitedefaultendpunct}{\mcitedefaultseppunct}\relax
\EndOfBibitem
\bibitem[Smith \latin{et~al.}(2019)Smith, Nebgen, Zubatyuk, Lubbers, Devereux, Barros, Tretiak, Isayev, and Roitberg]{Smith2019approaching}
Smith,~J.~S.; Nebgen,~B.~T.; Zubatyuk,~R.; Lubbers,~N.; Devereux,~C.; Barros,~K.; Tretiak,~S.; Isayev,~O.; Roitberg,~A.~E. \emph{Nat. Commun.} \textbf{2019}, \emph{10}, 2903\relax
\mciteBstWouldAddEndPuncttrue
\mciteSetBstMidEndSepPunct{\mcitedefaultmidpunct}
{\mcitedefaultendpunct}{\mcitedefaultseppunct}\relax
\EndOfBibitem
\bibitem[Gilmer \latin{et~al.}(2017)Gilmer, Schoenholz, Riley, Vinyals, and Dahl]{Gilmer2017neural}
Gilmer,~J.; Schoenholz,~S.~S.; Riley,~P.~F.; Vinyals,~O.; Dahl,~G.~E. Neural Message Passing for Quantum Chemistry. Proceedings of the 34th International Conference on Machine Learning. Sydney, Australia, 2017; pp 1263--1272\relax
\mciteBstWouldAddEndPuncttrue
\mciteSetBstMidEndSepPunct{\mcitedefaultmidpunct}
{\mcitedefaultendpunct}{\mcitedefaultseppunct}\relax
\EndOfBibitem
\bibitem[Sch{\"u}tt \latin{et~al.}(2017)Sch{\"u}tt, Arbabzadah, Chmiela, M{\"u}ller, and Tkatchenko]{schutt2017quantum}
Sch{\"u}tt,~K.~T.; Arbabzadah,~F.; Chmiela,~S.; M{\"u}ller,~K.~R.; Tkatchenko,~A. \emph{Nat. Commun.} \textbf{2017}, \emph{8}, 13890\relax
\mciteBstWouldAddEndPuncttrue
\mciteSetBstMidEndSepPunct{\mcitedefaultmidpunct}
{\mcitedefaultendpunct}{\mcitedefaultseppunct}\relax
\EndOfBibitem
\bibitem[Schütt \latin{et~al.}(2018)Schütt, Sauceda, Kindermans, Tkatchenko, and Müller]{Schutt2018}
Schütt,~K.~T.; Sauceda,~H.~E.; Kindermans,~P.-J.; Tkatchenko,~A.; Müller,~K.-R. \emph{J. Chem. Phys.} \textbf{2018}, \emph{148}, 241722\relax
\mciteBstWouldAddEndPuncttrue
\mciteSetBstMidEndSepPunct{\mcitedefaultmidpunct}
{\mcitedefaultendpunct}{\mcitedefaultseppunct}\relax
\EndOfBibitem
\bibitem[Unke and Meuwly(2019)Unke, and Meuwly]{unke2019physnet}
Unke,~O.~T.; Meuwly,~M. \emph{J. Chem. Theory Comput.} \textbf{2019}, \emph{15}, 3678--3693\relax
\mciteBstWouldAddEndPuncttrue
\mciteSetBstMidEndSepPunct{\mcitedefaultmidpunct}
{\mcitedefaultendpunct}{\mcitedefaultseppunct}\relax
\EndOfBibitem
\bibitem[Gasteiger \latin{et~al.}(2020)Gasteiger, Gro{\ss}, and G{\"u}nnemann]{gasteiger2020directional}
Gasteiger,~J.; Gro{\ss},~J.; G{\"u}nnemann,~S. \emph{arXiv preprint arXiv:2003.03123} \textbf{2020}, \relax
\mciteBstWouldAddEndPunctfalse
\mciteSetBstMidEndSepPunct{\mcitedefaultmidpunct}
{}{\mcitedefaultseppunct}\relax
\EndOfBibitem
\bibitem[Batzner \latin{et~al.}(2022)Batzner, Musaelian, Sun, Geiger, Mailoa, Kornbluth, Molinari, Smidt, and Kozinsky]{batzner20223}
Batzner,~S.; Musaelian,~A.; Sun,~L.; Geiger,~M.; Mailoa,~J.~P.; Kornbluth,~M.; Molinari,~N.; Smidt,~T.~E.; Kozinsky,~B. \emph{Nat. Commun.} \textbf{2022}, \emph{13}, 2453\relax
\mciteBstWouldAddEndPuncttrue
\mciteSetBstMidEndSepPunct{\mcitedefaultmidpunct}
{\mcitedefaultendpunct}{\mcitedefaultseppunct}\relax
\EndOfBibitem
\bibitem[Musaelian \latin{et~al.}(2023)Musaelian, Batzner, Johansson, Sun, Owen, Kornbluth, and Kozinsky]{musaelian2023learning}
Musaelian,~A.; Batzner,~S.; Johansson,~A.; Sun,~L.; Owen,~C.~J.; Kornbluth,~M.; Kozinsky,~B. \emph{Nat. Commun.} \textbf{2023}, \emph{14}, 579\relax
\mciteBstWouldAddEndPuncttrue
\mciteSetBstMidEndSepPunct{\mcitedefaultmidpunct}
{\mcitedefaultendpunct}{\mcitedefaultseppunct}\relax
\EndOfBibitem
\bibitem[Ramakrishnan \latin{et~al.}(2014)Ramakrishnan, Dral, Rupp, and Lilienfeld]{Ramakrishnan2014}
Ramakrishnan,~R.; Dral,~P.~O.; Rupp,~M.; Lilienfeld,~O. A.~V. \emph{Sci. Data} \textbf{2014}, \emph{1}, 1--7\relax
\mciteBstWouldAddEndPuncttrue
\mciteSetBstMidEndSepPunct{\mcitedefaultmidpunct}
{\mcitedefaultendpunct}{\mcitedefaultseppunct}\relax
\EndOfBibitem
\bibitem[Smith \latin{et~al.}(2017)Smith, Isayev, and Roitberg]{Smith2017dataset}
Smith,~J.~S.; Isayev,~O.; Roitberg,~A.~E. \emph{Sci. Data} \textbf{2017}, \emph{4}, 1--8\relax
\mciteBstWouldAddEndPuncttrue
\mciteSetBstMidEndSepPunct{\mcitedefaultmidpunct}
{\mcitedefaultendpunct}{\mcitedefaultseppunct}\relax
\EndOfBibitem
\bibitem[No{\'e} \latin{et~al.}(2020)No{\'e}, Tkatchenko, M{\"u}ller, and Clementi]{noe2020machine}
No{\'e},~F.; Tkatchenko,~A.; M{\"u}ller,~K.-R.; Clementi,~C. \emph{Annu. Rev. Phys. Chem.} \textbf{2020}, \emph{71}, 361--390\relax
\mciteBstWouldAddEndPuncttrue
\mciteSetBstMidEndSepPunct{\mcitedefaultmidpunct}
{\mcitedefaultendpunct}{\mcitedefaultseppunct}\relax
\EndOfBibitem
\bibitem[Tkachenko \latin{et~al.}(2023)Tkachenko, Tkachenko, Nebgen, Tretiak, and Boldyrev]{tkachenko2023neural}
Tkachenko,~N.~V.; Tkachenko,~A.~A.; Nebgen,~B.; Tretiak,~S.; Boldyrev,~A.~I. \emph{Phys. Chem. Chem. Phys.} \textbf{2023}, \emph{25}, 21173--21182\relax
\mciteBstWouldAddEndPuncttrue
\mciteSetBstMidEndSepPunct{\mcitedefaultmidpunct}
{\mcitedefaultendpunct}{\mcitedefaultseppunct}\relax
\EndOfBibitem
\bibitem[Kulichenko \latin{et~al.}(2021)Kulichenko, Smith, Nebgen, Li, Fedik, Boldyrev, Lubbers, Barros, and Tretiak]{kulichenko2021rise}
Kulichenko,~M.; Smith,~J.~S.; Nebgen,~B.; Li,~Y.~W.; Fedik,~N.; Boldyrev,~A.~I.; Lubbers,~N.; Barros,~K.; Tretiak,~S. \emph{J. Phys. Chem. Lett.} \textbf{2021}, \emph{12}, 6227--6243\relax
\mciteBstWouldAddEndPuncttrue
\mciteSetBstMidEndSepPunct{\mcitedefaultmidpunct}
{\mcitedefaultendpunct}{\mcitedefaultseppunct}\relax
\EndOfBibitem
\bibitem[Lan \latin{et~al.}(2023)Lan, Palizhati, Shuaibi, Wood, Wander, Das, Uyttendaele, Zitnick, and Ulissi]{lan2023adsorbml}
Lan,~J.; Palizhati,~A.; Shuaibi,~M.; Wood,~B.~M.; Wander,~B.; Das,~A.; Uyttendaele,~M.; Zitnick,~C.~L.; Ulissi,~Z.~W. \emph{npj Comput. Mater.} \textbf{2023}, \emph{9}, 172\relax
\mciteBstWouldAddEndPuncttrue
\mciteSetBstMidEndSepPunct{\mcitedefaultmidpunct}
{\mcitedefaultendpunct}{\mcitedefaultseppunct}\relax
\EndOfBibitem
\bibitem[Schreiner \latin{et~al.}(2022)Schreiner, Bhowmik, Vegge, Busk, and Winther]{Schreiner2022transition1x}
Schreiner,~M.; Bhowmik,~A.; Vegge,~T.; Busk,~J.; Winther,~O. \emph{Sci. Data} \textbf{2022}, \emph{9}, 779\relax
\mciteBstWouldAddEndPuncttrue
\mciteSetBstMidEndSepPunct{\mcitedefaultmidpunct}
{\mcitedefaultendpunct}{\mcitedefaultseppunct}\relax
\EndOfBibitem
\bibitem[Zhang \latin{et~al.}(2024)Zhang, Mako{\'s}, Jadrich, Kraka, Barros, Nebgen, Tretiak, Isayev, Lubbers, Messerly, \latin{et~al.} others]{zhang2024exploring}
Zhang,~S.; Mako{\'s},~M.~Z.; Jadrich,~R.~B.; Kraka,~E.; Barros,~K.; Nebgen,~B.~T.; Tretiak,~S.; Isayev,~O.; Lubbers,~N.; Messerly,~R.~A.; others \emph{Nat. Chem.} \textbf{2024}, \emph{16}, 727--734\relax
\mciteBstWouldAddEndPuncttrue
\mciteSetBstMidEndSepPunct{\mcitedefaultmidpunct}
{\mcitedefaultendpunct}{\mcitedefaultseppunct}\relax
\EndOfBibitem
\bibitem[Smith \latin{et~al.}(2018)Smith, Nebgen, Lubbers, Isayev, and Roitberg]{Smith2018less}
Smith,~J.~S.; Nebgen,~B.; Lubbers,~N.; Isayev,~O.; Roitberg,~A.~E. \emph{J. Chem. Phys.} \textbf{2018}, \emph{148}\relax
\mciteBstWouldAddEndPuncttrue
\mciteSetBstMidEndSepPunct{\mcitedefaultmidpunct}
{\mcitedefaultendpunct}{\mcitedefaultseppunct}\relax
\EndOfBibitem
\bibitem[Vandermause \latin{et~al.}(2020)Vandermause, Torrisi, Batzner, Xie, Sun, Kolpak, and Kozinsky]{vandermause2020fly}
Vandermause,~J.; Torrisi,~S.~B.; Batzner,~S.; Xie,~Y.; Sun,~L.; Kolpak,~A.~M.; Kozinsky,~B. \emph{npj Comput. Mater.} \textbf{2020}, \emph{6}, 20\relax
\mciteBstWouldAddEndPuncttrue
\mciteSetBstMidEndSepPunct{\mcitedefaultmidpunct}
{\mcitedefaultendpunct}{\mcitedefaultseppunct}\relax
\EndOfBibitem
\bibitem[Smith \latin{et~al.}(2021)Smith, Nebgen, Mathew, Chen, Lubbers, Burakovsky, Tretiak, Nam, Germann, Fensin, \latin{et~al.} others]{Smith2021automated}
Smith,~J.~S.; Nebgen,~B.; Mathew,~N.; Chen,~J.; Lubbers,~N.; Burakovsky,~L.; Tretiak,~S.; Nam,~H.~A.; Germann,~T.; Fensin,~S.; others \emph{Nat. Commun.} \textbf{2021}, \emph{12}, 1257\relax
\mciteBstWouldAddEndPuncttrue
\mciteSetBstMidEndSepPunct{\mcitedefaultmidpunct}
{\mcitedefaultendpunct}{\mcitedefaultseppunct}\relax
\EndOfBibitem
\bibitem[Kulichenko \latin{et~al.}(2023)Kulichenko, Barros, Lubbers, Li, Messerly, Tretiak, Smith, and Nebgen]{kulichenko2023uncertainty}
Kulichenko,~M.; Barros,~K.; Lubbers,~N.; Li,~Y.~W.; Messerly,~R.; Tretiak,~S.; Smith,~J.~S.; Nebgen,~B. \emph{Nat. Comput. Sci.} \textbf{2023}, \emph{3}, 230--239\relax
\mciteBstWouldAddEndPuncttrue
\mciteSetBstMidEndSepPunct{\mcitedefaultmidpunct}
{\mcitedefaultendpunct}{\mcitedefaultseppunct}\relax
\EndOfBibitem
\bibitem[Grambow \latin{et~al.}(2020)Grambow, Pattanaik, and Green]{Grambow2020dataset}
Grambow,~C.~A.; Pattanaik,~L.; Green,~W.~H. \emph{Sci. Data} \textbf{2020}, \emph{7}, 1--8\relax
\mciteBstWouldAddEndPuncttrue
\mciteSetBstMidEndSepPunct{\mcitedefaultmidpunct}
{\mcitedefaultendpunct}{\mcitedefaultseppunct}\relax
\EndOfBibitem
\bibitem[van Gerwen \latin{et~al.}(2024)van Gerwen, Briling, Bunne, Somnath, Laplaza, Krause, and Corminboeuf]{van20243dreact}
van Gerwen,~P.; Briling,~K.~R.; Bunne,~C.; Somnath,~V.~R.; Laplaza,~R.; Krause,~A.; Corminboeuf,~C. \emph{J. Chem. Inf. Model.} \textbf{2024}, \relax
\mciteBstWouldAddEndPunctfalse
\mciteSetBstMidEndSepPunct{\mcitedefaultmidpunct}
{}{\mcitedefaultseppunct}\relax
\EndOfBibitem
\bibitem[Spiekermann \latin{et~al.}(2022)Spiekermann, Pattanaik, and Green]{Spiekermann2022}
Spiekermann,~K.~A.; Pattanaik,~L.; Green,~W.~H. \emph{J. Phys. Chem.} \textbf{2022}, \emph{126}, 3976--3986\relax
\mciteBstWouldAddEndPuncttrue
\mciteSetBstMidEndSepPunct{\mcitedefaultmidpunct}
{\mcitedefaultendpunct}{\mcitedefaultseppunct}\relax
\EndOfBibitem
\bibitem[Heinen \latin{et~al.}(2021)Heinen, von Rudorff, and von Lilienfeld]{Heinen2021}
Heinen,~S.; von Rudorff,~G.~F.; von Lilienfeld,~O.~A. \emph{J. Chem. Phys.} \textbf{2021}, \emph{155}, 064105\relax
\mciteBstWouldAddEndPuncttrue
\mciteSetBstMidEndSepPunct{\mcitedefaultmidpunct}
{\mcitedefaultendpunct}{\mcitedefaultseppunct}\relax
\EndOfBibitem
\bibitem[Grambow \latin{et~al.}(2020)Grambow, Pattanaik, and Green]{grambow2020deep}
Grambow,~C.~A.; Pattanaik,~L.; Green,~W.~H. \emph{J. Phys. Chem. Lett.} \textbf{2020}, \emph{11}, 2992--2997\relax
\mciteBstWouldAddEndPuncttrue
\mciteSetBstMidEndSepPunct{\mcitedefaultmidpunct}
{\mcitedefaultendpunct}{\mcitedefaultseppunct}\relax
\EndOfBibitem
\bibitem[Makos \latin{et~al.}(2021)Makos, Verma, and Larson]{Makos2021}
Makos,~M.~Z.; Verma,~N.; Larson,~E.~C. \emph{J. Chem. Phys.} \textbf{2021}, 024116\relax
\mciteBstWouldAddEndPuncttrue
\mciteSetBstMidEndSepPunct{\mcitedefaultmidpunct}
{\mcitedefaultendpunct}{\mcitedefaultseppunct}\relax
\EndOfBibitem
\bibitem[Jackson \latin{et~al.}(2021)Jackson, Zhang, and Pearson]{Jackson2021}
Jackson,~R.; Zhang,~W.; Pearson,~J. \emph{Chem. Sci.} \textbf{2021}, \emph{12}, 10022--10040\relax
\mciteBstWouldAddEndPuncttrue
\mciteSetBstMidEndSepPunct{\mcitedefaultmidpunct}
{\mcitedefaultendpunct}{\mcitedefaultseppunct}\relax
\EndOfBibitem
\bibitem[Pattanaik \latin{et~al.}(2020)Pattanaik, Ingraham, Grambow, and Green]{Pattanaik2020}
Pattanaik,~L.; Ingraham,~J.~B.; Grambow,~C.~A.; Green,~W.~H. \emph{Phys. Chem. Chem. Phys.} \textbf{2020}, \emph{22}, 23618--23626\relax
\mciteBstWouldAddEndPuncttrue
\mciteSetBstMidEndSepPunct{\mcitedefaultmidpunct}
{\mcitedefaultendpunct}{\mcitedefaultseppunct}\relax
\EndOfBibitem
\bibitem[Stocker \latin{et~al.}(2020)Stocker, Csanyi, Reuter, and Margraf]{stocker2020machine}
Stocker,~S.; Csanyi,~G.; Reuter,~K.; Margraf,~J.~T. \emph{Nat. Commun.} \textbf{2020}, \emph{11}, 5505\relax
\mciteBstWouldAddEndPuncttrue
\mciteSetBstMidEndSepPunct{\mcitedefaultmidpunct}
{\mcitedefaultendpunct}{\mcitedefaultseppunct}\relax
\EndOfBibitem
\bibitem[Meuwly(2021)]{meuwly2021machine}
Meuwly,~M. \emph{Chem. Rev.} \textbf{2021}, \emph{121}, 10218--10239\relax
\mciteBstWouldAddEndPuncttrue
\mciteSetBstMidEndSepPunct{\mcitedefaultmidpunct}
{\mcitedefaultendpunct}{\mcitedefaultseppunct}\relax
\EndOfBibitem
\bibitem[van Gerwen \latin{et~al.}(2022)van Gerwen, Fabrizio, Wodrich, and Corminboeuf]{vanGerwen2022physics}
van Gerwen,~P.; Fabrizio,~A.; Wodrich,~M.~D.; Corminboeuf,~C. \emph{Mach. Learn.: Sci. Technol.} \textbf{2022}, \emph{3}, 045005\relax
\mciteBstWouldAddEndPuncttrue
\mciteSetBstMidEndSepPunct{\mcitedefaultmidpunct}
{\mcitedefaultendpunct}{\mcitedefaultseppunct}\relax
\EndOfBibitem
\bibitem[Heid and Green(2021)Heid, and Green]{Heid2021}
Heid,~E.; Green,~W.~H. \emph{J. Chem. Inf. Model.} \textbf{2021}, \emph{62}, 2101--2110\relax
\mciteBstWouldAddEndPuncttrue
\mciteSetBstMidEndSepPunct{\mcitedefaultmidpunct}
{\mcitedefaultendpunct}{\mcitedefaultseppunct}\relax
\EndOfBibitem
\bibitem[Baker(1987)]{Baker1987}
Baker,~J. \emph{J. Comput. Chem.} \textbf{1987}, \emph{8}, 563--574\relax
\mciteBstWouldAddEndPuncttrue
\mciteSetBstMidEndSepPunct{\mcitedefaultmidpunct}
{\mcitedefaultendpunct}{\mcitedefaultseppunct}\relax
\EndOfBibitem
\bibitem[Peng \latin{et~al.}(1996)Peng, Ayala, Schlegel, and Frisch]{peng1996using}
Peng,~C.; Ayala,~P.~Y.; Schlegel,~H.~B.; Frisch,~M.~J. \emph{J. Comput. Chem.} \textbf{1996}, \emph{17}, 49--56\relax
\mciteBstWouldAddEndPuncttrue
\mciteSetBstMidEndSepPunct{\mcitedefaultmidpunct}
{\mcitedefaultendpunct}{\mcitedefaultseppunct}\relax
\EndOfBibitem
\bibitem[Bofill(1994)]{bofill1994updated}
Bofill,~J.~M. \emph{J. Comput. Chem.} \textbf{1994}, \emph{15}, 1--11\relax
\mciteBstWouldAddEndPuncttrue
\mciteSetBstMidEndSepPunct{\mcitedefaultmidpunct}
{\mcitedefaultendpunct}{\mcitedefaultseppunct}\relax
\EndOfBibitem
\bibitem[Henkelman and J{\'o}nsson(1999)Henkelman, and J{\'o}nsson]{henkelman1999dimer}
Henkelman,~G.; J{\'o}nsson,~H. \emph{J. Chem. Phys.} \textbf{1999}, \emph{111}, 7010--7022\relax
\mciteBstWouldAddEndPuncttrue
\mciteSetBstMidEndSepPunct{\mcitedefaultmidpunct}
{\mcitedefaultendpunct}{\mcitedefaultseppunct}\relax
\EndOfBibitem
\bibitem[Sharada \latin{et~al.}(2012)Sharada, Zimmerman, Bell, and Head-Gordon]{mallikarjun2012automated}
Sharada,~S.~M.; Zimmerman,~P.~M.; Bell,~A.~T.; Head-Gordon,~M. \emph{J. Chem. Theory Comput.} \textbf{2012}, \emph{8}, 5166--5174\relax
\mciteBstWouldAddEndPuncttrue
\mciteSetBstMidEndSepPunct{\mcitedefaultmidpunct}
{\mcitedefaultendpunct}{\mcitedefaultseppunct}\relax
\EndOfBibitem
\bibitem[Schlegel(1982)]{schlegel1982optimization}
Schlegel,~H.~B. \emph{J. Comput. Chem.} \textbf{1982}, \emph{3}, 214--218\relax
\mciteBstWouldAddEndPuncttrue
\mciteSetBstMidEndSepPunct{\mcitedefaultmidpunct}
{\mcitedefaultendpunct}{\mcitedefaultseppunct}\relax
\EndOfBibitem
\bibitem[Bart{\'o}k \latin{et~al.}(2010)Bart{\'o}k, Payne, Kondor, and Cs{\'a}nyi]{bartok2010gaussian}
Bart{\'o}k,~A.~P.; Payne,~M.~C.; Kondor,~R.; Cs{\'a}nyi,~G. \emph{Phys. Rev. Lett.} \textbf{2010}, \emph{104}, 136403\relax
\mciteBstWouldAddEndPuncttrue
\mciteSetBstMidEndSepPunct{\mcitedefaultmidpunct}
{\mcitedefaultendpunct}{\mcitedefaultseppunct}\relax
\EndOfBibitem
\bibitem[Denzel \latin{et~al.}(2019)Denzel, Haasdonk, and Kästner]{denzel2019gaussian}
Denzel,~A.; Haasdonk,~B.; Kästner,~J. \emph{J. Phys. Chem. A} \textbf{2019}, \emph{123}, 9600--9611\relax
\mciteBstWouldAddEndPuncttrue
\mciteSetBstMidEndSepPunct{\mcitedefaultmidpunct}
{\mcitedefaultendpunct}{\mcitedefaultseppunct}\relax
\EndOfBibitem
\bibitem[Denzel and Kästner(2020)Denzel, and Kästner]{denzel2020hessian}
Denzel,~A.; Kästner,~J. \emph{J. Chem. Theory Comput.} \textbf{2020}, \emph{16}, 5083--5089\relax
\mciteBstWouldAddEndPuncttrue
\mciteSetBstMidEndSepPunct{\mcitedefaultmidpunct}
{\mcitedefaultendpunct}{\mcitedefaultseppunct}\relax
\EndOfBibitem
\bibitem[Xie \latin{et~al.}(2021)Xie, Vandermause, Sun, Cepellotti, and Kozinsky]{xie2021bayesian}
Xie,~Y.; Vandermause,~J.; Sun,~L.; Cepellotti,~A.; Kozinsky,~B. \emph{npj Comput. Mater.} \textbf{2021}, \emph{7}, 40\relax
\mciteBstWouldAddEndPuncttrue
\mciteSetBstMidEndSepPunct{\mcitedefaultmidpunct}
{\mcitedefaultendpunct}{\mcitedefaultseppunct}\relax
\EndOfBibitem
\bibitem[Schreiner \latin{et~al.}(2022)Schreiner, Bhowmik, Vegge, Jørgensen, and Winther]{Schreiner2022NNEB}
Schreiner,~M.; Bhowmik,~A.; Vegge,~T.; Jørgensen,~P.~B.; Winther,~O. \emph{Mach. Learn.: Sci. Technol.} \textbf{2022}, \emph{3}, 045022\relax
\mciteBstWouldAddEndPuncttrue
\mciteSetBstMidEndSepPunct{\mcitedefaultmidpunct}
{\mcitedefaultendpunct}{\mcitedefaultseppunct}\relax
\EndOfBibitem
\bibitem[Wander \latin{et~al.}(2024)Wander, Shuaibi, Kitchin, Ulissi, and Zitnick]{wander2024cattsunami}
Wander,~B.; Shuaibi,~M.; Kitchin,~J.~R.; Ulissi,~Z.~W.; Zitnick,~C.~L. \emph{arXiv preprint arXiv:2405.02078} \textbf{2024}, \relax
\mciteBstWouldAddEndPunctfalse
\mciteSetBstMidEndSepPunct{\mcitedefaultmidpunct}
{}{\mcitedefaultseppunct}\relax
\EndOfBibitem
\bibitem[Fink \latin{et~al.}(2005)Fink, Bruggesser, and Reymond]{Fink2005}
Fink,~T.; Bruggesser,~H.; Reymond,~J.-L. \emph{Angew. Chem. Int. Ed.} \textbf{2005}, \emph{44}, 1504--1508\relax
\mciteBstWouldAddEndPuncttrue
\mciteSetBstMidEndSepPunct{\mcitedefaultmidpunct}
{\mcitedefaultendpunct}{\mcitedefaultseppunct}\relax
\EndOfBibitem
\bibitem[Fink and Reymond(2007)Fink, and Reymond]{Fink2007}
Fink,~T.; Reymond,~J.-L. \emph{J. Chem. Inf. Model.} \textbf{2007}, \emph{47}, 342--353\relax
\mciteBstWouldAddEndPuncttrue
\mciteSetBstMidEndSepPunct{\mcitedefaultmidpunct}
{\mcitedefaultendpunct}{\mcitedefaultseppunct}\relax
\EndOfBibitem
\bibitem[Fey and Lenssen(2019)Fey, and Lenssen]{fey2019fast}
Fey,~M.; Lenssen,~J.~E. \emph{arXiv preprint arXiv:1903.02428} \textbf{2019}, \relax
\mciteBstWouldAddEndPunctfalse
\mciteSetBstMidEndSepPunct{\mcitedefaultmidpunct}
{}{\mcitedefaultseppunct}\relax
\EndOfBibitem
\bibitem[Snoek \latin{et~al.}(2012)Snoek, Larochelle, and Adams]{Snoek2012}
Snoek,~J.; Larochelle,~H.; Adams,~R.~P. Practical Bayesian Optimization of Machine Learning Algorithms. Advances in Neural Information Processing Systems 25. Lake Tahoe, Nevada, USA, 2012; pp 2951--2959\relax
\mciteBstWouldAddEndPuncttrue
\mciteSetBstMidEndSepPunct{\mcitedefaultmidpunct}
{\mcitedefaultendpunct}{\mcitedefaultseppunct}\relax
\EndOfBibitem
\bibitem[Bakshy \latin{et~al.}(2018)Bakshy, Dworkin, Karrer, Kashin, Letham, Murthy, and Singh]{Bakshy2018}
Bakshy,~E.; Dworkin,~L.; Karrer,~B.; Kashin,~K.; Letham,~B.; Murthy,~A.; Singh,~S. AE: A Domain-Agnostic Platform for Adaptive Experimentation. Advances in Neural Information Processing Systems 31. Montr{\'e}al, Canada, 2018; pp 1--8\relax
\mciteBstWouldAddEndPuncttrue
\mciteSetBstMidEndSepPunct{\mcitedefaultmidpunct}
{\mcitedefaultendpunct}{\mcitedefaultseppunct}\relax
\EndOfBibitem
\bibitem[Chai and Head-Gordon(2008)Chai, and Head-Gordon]{Chai2008}
Chai,~J.-D.; Head-Gordon,~M. \emph{J. Chem. Phys.} \textbf{2008}, \emph{128}\relax
\mciteBstWouldAddEndPuncttrue
\mciteSetBstMidEndSepPunct{\mcitedefaultmidpunct}
{\mcitedefaultendpunct}{\mcitedefaultseppunct}\relax
\EndOfBibitem
\bibitem[Ditchfield \latin{et~al.}(1971)Ditchfield, Hehre, and Pople]{Ditchfield1971}
Ditchfield,~R.; Hehre,~W.~J.; Pople,~J.~A. \emph{J. Chem. Phys.} \textbf{1971}, \emph{54}, 724--728\relax
\mciteBstWouldAddEndPuncttrue
\mciteSetBstMidEndSepPunct{\mcitedefaultmidpunct}
{\mcitedefaultendpunct}{\mcitedefaultseppunct}\relax
\EndOfBibitem
\bibitem[Gill \latin{et~al.}(1993)Gill, Johnson, and Pople]{gill1993standard}
Gill,~P.~M.; Johnson,~B.~G.; Pople,~J.~A. \emph{Chem. Phys. Lett.} \textbf{1993}, \emph{209}, 506--512\relax
\mciteBstWouldAddEndPuncttrue
\mciteSetBstMidEndSepPunct{\mcitedefaultmidpunct}
{\mcitedefaultendpunct}{\mcitedefaultseppunct}\relax
\EndOfBibitem
\bibitem[Epifanovsky \latin{et~al.}(2021)Epifanovsky, Gilbert, Feng, Lee, Mao, Mardirossian, Pokhilko, White, Coons, Dempwolff, \latin{et~al.} others]{Epifanovsky2021}
Epifanovsky,~E.; Gilbert,~A.~T.; Feng,~X.; Lee,~J.; Mao,~Y.; Mardirossian,~N.; Pokhilko,~P.; White,~A.~F.; Coons,~M.~P.; Dempwolff,~A.~L.; others \emph{J. Chem. Phys.} \textbf{2021}, \emph{155}, 084801\relax
\mciteBstWouldAddEndPuncttrue
\mciteSetBstMidEndSepPunct{\mcitedefaultmidpunct}
{\mcitedefaultendpunct}{\mcitedefaultseppunct}\relax
\EndOfBibitem
\bibitem[Schmidt \latin{et~al.}(1985)Schmidt, Gordon, and Dupuis]{schmidt1985intrinsic}
Schmidt,~M.~W.; Gordon,~M.~S.; Dupuis,~M. \emph{J. Am. Chem. Soc.} \textbf{1985}, \emph{107}, 2585--2589\relax
\mciteBstWouldAddEndPuncttrue
\mciteSetBstMidEndSepPunct{\mcitedefaultmidpunct}
{\mcitedefaultendpunct}{\mcitedefaultseppunct}\relax
\EndOfBibitem
\bibitem[Blanchard~Jr and Cairncross(1966)Blanchard~Jr, and Cairncross]{blanchard1966bicyclo}
Blanchard~Jr,~E.~P.; Cairncross,~A. \emph{J. Am. Chem. Soc.} \textbf{1966}, \emph{88}, 487--495\relax
\mciteBstWouldAddEndPuncttrue
\mciteSetBstMidEndSepPunct{\mcitedefaultmidpunct}
{\mcitedefaultendpunct}{\mcitedefaultseppunct}\relax
\EndOfBibitem
\bibitem[Closs and Pfeffer(1968)Closs, and Pfeffer]{closs1968steric}
Closs,~G.; Pfeffer,~P. \emph{J. Am. Chem. Soc.} \textbf{1968}, \emph{90}, 2452--2453\relax
\mciteBstWouldAddEndPuncttrue
\mciteSetBstMidEndSepPunct{\mcitedefaultmidpunct}
{\mcitedefaultendpunct}{\mcitedefaultseppunct}\relax
\EndOfBibitem
\bibitem[Dewar and Kirschner(1975)Dewar, and Kirschner]{dewar1975mindo}
Dewar,~M.~J.; Kirschner,~S. \emph{J. Am. Chem. Soc.} \textbf{1975}, \emph{97}, 2931--2932\relax
\mciteBstWouldAddEndPuncttrue
\mciteSetBstMidEndSepPunct{\mcitedefaultmidpunct}
{\mcitedefaultendpunct}{\mcitedefaultseppunct}\relax
\EndOfBibitem
\bibitem[Shevlin and McKee(1988)Shevlin, and McKee]{shevlin1988theoretical}
Shevlin,~P.~B.; McKee,~M.~L. \emph{J. Am. Chem. Soc.} \textbf{1988}, \emph{110}, 1666--1671\relax
\mciteBstWouldAddEndPuncttrue
\mciteSetBstMidEndSepPunct{\mcitedefaultmidpunct}
{\mcitedefaultendpunct}{\mcitedefaultseppunct}\relax
\EndOfBibitem
\bibitem[Mironov \latin{et~al.}(2019)Mironov, Alexeev, Mulligan, and Fedorov]{mironov2019systematic}
Mironov,~V.; Alexeev,~Y.; Mulligan,~V.~K.; Fedorov,~D.~G. \emph{J. Comput. Chem.} \textbf{2019}, \emph{40}, 297--309\relax
\mciteBstWouldAddEndPuncttrue
\mciteSetBstMidEndSepPunct{\mcitedefaultmidpunct}
{\mcitedefaultendpunct}{\mcitedefaultseppunct}\relax
\EndOfBibitem
\bibitem[Vymetal and Vondrasek(2010)Vymetal, and Vondrasek]{vymetal2010metadynamics}
Vymetal,~J.; Vondrasek,~J. \emph{J. Phys. Chem. B.} \textbf{2010}, \emph{114}, 5632--5642\relax
\mciteBstWouldAddEndPuncttrue
\mciteSetBstMidEndSepPunct{\mcitedefaultmidpunct}
{\mcitedefaultendpunct}{\mcitedefaultseppunct}\relax
\EndOfBibitem
\bibitem[Lubbers \latin{et~al.}(2018)Lubbers, Smith, and Barros]{lubbers2018hierarchical}
Lubbers,~N.; Smith,~J.~S.; Barros,~K. \emph{J. Chem. Phys.} \textbf{2018}, \emph{148}\relax
\mciteBstWouldAddEndPuncttrue
\mciteSetBstMidEndSepPunct{\mcitedefaultmidpunct}
{\mcitedefaultendpunct}{\mcitedefaultseppunct}\relax
\EndOfBibitem
\end{mcitethebibliography}
\end{document}